\documentclass[utf8]{FrontiersinHarvard} 

\usepackage{url,hyperref,lineno}
\usepackage[nopatch=footnote]{microtype}
\newcommand*{\doi}[1]{\begingroup\urlstyle{same}\href{https://doi.org/#1}{doi:\discretionary{}{}{}\nolinkurl{#1}}\endgroup}
\usepackage[onehalfspacing]{setspace}

\def\keyFont{\fontsize{8}{11}\helveticabold }
\def\firstAuthorLast{Cao {et~al.}} 
\def\Authors{Wenxue Cao\,$^{1,*}$, Kristoffer Aalstad\,$^{1}$, Louise S. Schmidt\,$^{1}$ and Thomas V. Schuler\,$^{1}$}
\def\Address{$^{1}$Department of Geosciences, University of Oslo, Oslo, Norway
}
\def\corrAuthor{Wenxue Cao}

\def\corrEmail{wenxue.cao@geo.uio.no}

\begin{document}
\onecolumn
\firstpage{1}

\title[Cubic]{Quantity, quality, and timing: Guiding glacier data assimilation strategies in the high Arctic}

\author[\firstAuthorLast ]{\Authors} 
\address{} 
\correspondance{} 

\extraAuth{}

\maketitle
\begin{abstract}
\section{}
Accurate simulation of glacier surface mass balance is essential for predicting sea level rise and freshwater resources, but it is constrained by uncertainties in meteorological forcing and model parameters. Here, we deploy glacier data assimilation strategies to assess the value of observations for improving surface mass balance simulation, focusing on observation quantity, quality, and timing. We perform synthetic twin experiments on Kongsvegen glacier, Svalbard, using a Particle Batch Smoother with 1000 ensemble members. Synthetic observations of albedo, snow depth, and surface temperature are assimilated at two quality levels, under two climatic scenarios, and over 12 years. Assimilation benefit is measured as the percentage improvement in the continuous ranked probability score of the posterior glacier surface mass balance relative to the prior. A single optimally timed high quality observation yields mean improvements of up to 80\%. Larger numbers of low quality observations partially compensate for lower improvement. In the accumulation zone, however, additional snow depth observations degrade performance through particle degeneracy. Optimal timing is governed by the seasonal transitions of the truth trajectory rather than by prior ensemble spread alone. The optimal windows shift by up to six weeks between early and late melting years. Joint assimilation adds value through temporal diversity rather than observational diversity, while independently timed observations outperform same day combinations. The asynchronously optimally timed combined assimilation of three variables sustains improvements of 85 to 97\% across all years in the ablation zone. These findings provide guidelines for adaptive observation scheduling in glacier monitoring and reanalysis.

\tiny
 \keyFont{ \section{Keywords:} data assimilation strategies, glacier surface mass balance, Particle Batch Smoother} 
\end{abstract}

\section{Introduction}
Mountain and Arctic glaciers are important contributors to sea-level rise and regional freshwater resources, and their accelerating mass loss over recent decades has intensified the need for accurate and efficient monitoring and modelling systems \citep{GlaMBIE2025CommunityChanges,Zemp2019Global2016}. In the Svalbard archipelago, where glaciers cover approximately 32\,500\,km$^2$ \citep{Kohler2021SvalbardInventory}, the climatic mass balance has been predominantly negative since around 2000 \citep{Schuler2020ReconcilingBalance}. Observations and models consistently document these negative mass balances and project that they will intensify under continued atmospheric warming \citep{VanPelt2021AcceleratingEnsemble}. Accurate simulations of glacier surface mass balance are essential for projecting future contributions to sea-level rise and regional hydrology. At the same time, simulations are hampered by uncertainties in meteorological forcing, surface process parameterisation, and model parameters \citep{Marzeion2020PartitioningChange, Schuster2023GlacierStrategies}.

Several observations are particularly informative for constraining glacier surface mass balance models. Surface albedo governs the absorption of solar radiation during the melt season, and its seasonal evolution, from fresh snow with high albedo values, through reduced values for aged snow   to lower bare-ice values, is closely tied to the  surface energy balance and thus the annual mass balance \citep{Dumont2012LinkingData,Dumont2012VariationalGlacier, Moller2017ModelingHiRSvaC, Ye2024UnveilingGlaciers}. Snow depth, together with density, quantifies the mass stored in the winter snowpack and modulates the timing and magnitude of meltwater runoff \citep{VanPelt2019A1957-2018, An2020SnowReflectometry}. Surface temperature helps determine turbulent sensible and latent heat fluxes at the glacier surface and influences the partitioning of energy between melt, refreezing, and heat conduction into the ice \citep{stby2014SevereSvalbard}. Observations of these variables are becoming increasingly available through various satellite platforms, albeit with limitations in spatial resolution, temporal revisit frequency, and accuracy, particularly in high-latitude environments subject to persistent cloud cover \citep{Stroeve2005AccuracyMeasurements,Aalstad2020}.

Data assimilation (DA) provides a framework for combining uncertain model simulations with noisy observations, propagating observational information through the model state to improve simulation accuracy and reduce uncertainty \citep{Carrassi2018DataPerspectives, Evensen2022DataFundamentals}. Ensemble-based methods are particularly attractive for glacier applications because they accommodate the non-linear dynamics of surface energy and mass balance processes and naturally represent model uncertainty through the spread of ensemble realizations without the need for gradients \citep{Alonso-Gonzalez2022TheV1.0}. Among these, the Particle Batch Smoother (PBS) has demonstrated strong performance for seasonal snowpack estimation \citep{Margulis2015AEstimation, Margulis2016,Alonso-Gonzalez2022TheV1.0, Alonso-Gonzalez2023ExploringExperiment}, as well as for inferring the surface mass balance on Arctic glaciers \citep{Cao2025BayesianExperiments} and ice sheets \citep{Navari2021Reanalysis20002014}. The PBS is a so-called particle method\citep{Chopin2020} that weights ensemble members according to their consistency with observations over a trajectory through a Bayesian update.

Although ensemble-based DA has proven effective for glacier mass balance estimation \citep{Landmann2021AssimilatingFilter,Bertoncini2024AssimilationHydrology,Cao2025BayesianExperiments,Herrmann2025}, fundamental questions remain about how to design observational strategies to enhance performance. These include the selection of optimal observation variables, the identification of the most informative observation timings across the seasonal cycle, the quantification of trade-offs between observation quality and quantity, and the relative merits of assimilating single versus multiple variables  jointly \citep{Smyth2020ImprovingTiming,Guidicelli2024}. An improved understanding of when and what to observe, including whether a larger number of less precise observations can substitute for fewer high quality ones, is directly relevant to designing glacier data assimilation systems for practical reanalysis, monitoring, and forecasting. Satellite based systems are subject to revisit-time limitations and cloud contamination, while ground based campaigns are logistically constrained and cannot be operated continuously \citep{Kaser2003A2003}.

This study investigates these questions through a large ensemble of synthetic twin experiments on the Kongsvegen glacier, Svalbard, building on an established PBS framework for Arctic glacier surface mass balance \citep{Cao2025BayesianExperiments}. Our specific objectives are fourfold. We identify the most effective observation types and quantify how improvement converges with observation count. We assess the trade-off between observation quality and quantity across different observation types and glacier zones. We explain why certain observation timings are most effective by linking them to the structure of the synthetic truth trajectory and its seasonal transitions. We then compare individual and joint assimilation strategies to determine whether combining observation types provides additional benefits beyond the best individual observation assimilation strategies.

\section{Data and Methods}
\subsection{Study site and synthetic observations}
The study is conducted within a synthetic twin experiment framework on Kongsvegen glacier in Svalbard, Norway. The analysis is carried out separately for the ablation (ABL), equilibrium line altitude (ELA), and accumulation (ACC) zones of the glacier. All simulations use the glacier surface mass balance configuration of the CryoGrid community model \citep{Westermann2023CryoGridCommunity}, which solves the full surface energy balance and includes the snow and firn module first employed by \citet{Schmidt2023MeltwaterSvalbard}. The model is forced by the Copernicus Arctic Regional Reanalysis (CARRA), which provides air temperature, specific humidity, wind speed, incoming shortwave and longwave radiation, precipitation, and atmospheric pressure at a spatial resolution of 2.5\,km and a temporal resolution of 3\,hours \citep{C3S2024CARRA}.

The prior ensemble consists of $1000$ model runs generated by sampling two uncertain parameters from independent generalized logit-normal prior distributions, following \citet{Cao2025BayesianExperiments}. These two parameters were identified in \citet{Cao2025BayesianExperiments} as the most uncertain model inputs with the largest impact on the simulated surface mass balance. The albedo evolution rate $\tau_a$ is an inverse timescale that controls how rapidly the snow albedo decays with snow age in the Crocus albedo parameterisation \citep{Vionnet2012TheV7.2}. Its prior is bounded between $0.0001$ and $0.05$\,day$^{-1}$ with a median of $0.005$\,day$^{-1}$. The snowfall factor $\beta_s$ is a multiplicative correction of the CARRA snowfall, with a prior bounded between $0.5$ and $2$ and a median of $1$. This factor accounts for biases in the solid precipitation forcing and implicitly for unresolved redistribution of snow by wind. Both parameters are held fixed within each mass balance year, which defines the assimilation window. The same parameter prior is applied in every year, while each year's ensemble is driven by the meteorological forcing of that year. The resulting ensemble is designed to represent the plausible range of glacier states prior to assimilation.

The three observation types considered are surface albedo, snow depth, and surface temperature. These variables represent complementary constraints on the glacier surface energy and mass balance. Albedo governs the melt season radiative energy input through its control on shortwave radiation absorption and exhibits a characteristic seasonal transition from high snow albedo values (${\geq}0.8$) in winter to lower bare-ice albedo values (${\approx}0.4$) during the ablation season, with significant interannual variability in the timing and rate of this transition \citep{Dumont2012LinkingData, Davaze2018MonitoringData}. Snow depth is closely related to snow mass accumulated during winter, and indirectly controls the timing of the transition to bare ice during summer melt \citep{VanPelt2019A1957-2018, An2020SnowReflectometry}. Surface temperature is linked to the turbulent heat exchange at the glacier surface and to the thermal state of the snowpack, influencing both melt rates and refreezing during and after summer \citep{Karner2013ASvalbard}.

Synthetic observations are generated from the synthetic truth simulations introduced below by adding independent Gaussian noise with prescribed standard deviations. A synthetic observation of each variable is therefore available for every day of the simulation year. Two noise levels are considered for each variable, including high quality albedo ($\sigma_\alpha = 0.1$), low quality albedo ($\sigma_\alpha = 0.2$), high quality snow depth ($\sigma_d = 0.5$\,m), low quality snow depth ($\sigma_d = 1.0$\,m), high quality surface temperature ($\sigma_T = 1$\,K), and low quality surface temperature ($\sigma_T = 2$\,K). These noise levels bracket the range of uncertainties reported for satellite-derived and in situ observations of these variables in cold regions \citep{Stroeve2005AccuracyMeasurements, stby2014SevereSvalbard,Mazzolini2024Spatio-temporalAltimeter}.

Two contrasting climatic scenarios are employed to generate the synthetic truth. The two scenarios share the same CARRA forcing and the same prior ensemble, and they differ only in the true parameter values prescribed for the truth run. Scenario 1 (S1) combines a rapid albedo evolution rate with a high snowfall factor, using true values $\tau_a^\ast = 0.03$\,day$^{-1}$ and $\beta_s^\ast = 1.7$. This scenario generates pronounced seasonal signals, high interannual variability, and sharp melt season albedo transitions. Scenario 2 (S2) combines a slow albedo evolution rate with a low snowfall factor, using true values $\tau_a^\ast = 0.0003$\,day$^{-1}$ and $\beta_s^\ast = 0.7$. This scenario produces more gradual surface transitions, a reduced snowpack, and lower interannual variability. Both true parameter combinations lie in the tails of the prior distributions, which makes the assimilation exercise more challenging and more realistic than placing the truth near the prior median \citep{Cao2025BayesianExperiments}. Each scenario spans 12 simulation years (2010-2022). The interannual variability of the forcing produces both early and late melt seasons within each scenario.

\subsection{Data assimilation approach}
We employ the Particle Batch Smoother (PBS), a particle method originally developed for snow water equivalent estimation \citep{Margulis2015AEstimation} and subsequently applied to arctic snow cover and glacier surface mass balance \citep{Aalstad2018Ensemble-basedSites,Cao2025BayesianExperiments}. Like all particle methods, the PBS is a fully Bayesian method that requires no linearisation and requires no assumption of Gaussian prior or posterior distributions \citep{Chopin2020,Evensen2022DataFundamentals}.

Given an observation vector $\mathbf{y} = \left[y_1,\dots,y_k,\dots,y_N \right]$ containing observations at all assimilation times $t_k$ for $k\in1:N$ in the given assimilation window, the PBS assigns a particle weight to the model state trajectory $\mathbf{x}_i$ of each particle (ensemble member) $i$ with parameter vector $\boldsymbol{\theta}_i$ in the window according to the likelihood 
\begin{equation}
w_i  \propto p(\mathbf{y} \mid \boldsymbol{\theta}_i) \, ,
\label{eq:likelihood}
\end{equation}
by applying self-normalized importance sampling to Bayes' rule using the prior over parameters $\boldsymbol{\theta}$ as the proposal distribution $\boldsymbol{\theta}_i\sim p(\boldsymbol{\theta})$ \citep{Aalstad2026}.  Assuming conditionally independent observations, the likelihood factorizes as follows
\begin{equation}
    w_i\propto \prod_{k=1}^N p(y_k \mid \boldsymbol{\theta}_i) \, ,
\end{equation}
which, assuming typical additive Gaussian observation errors \citep{Carrassi2018DataPerspectives,Alonso-Gonzalez2022TheV1.0,Cao2025BayesianExperiments}, yields 
\begin{equation}
    w_i \propto \exp\left(-\frac{1}{2}\sum_{k=1}^N\frac{\left[y_k-\widehat{y}_k^{(i)}\right]^2}{2\sigma_k^2}\right)
\end{equation}
where $\widehat{y}_k^{(i)}=\widehat{y}_k(\boldsymbol{\theta}_i)$ denotes the (noise-free) predicted observation from the forward model (i.e., CryoGrid) with parameter vector $\boldsymbol{\theta}_i$ for particle $i$ and $\sigma_k$ is the observation error standard deviation of the $k$-th observation. The observations $y_k$ contained in the vector $\mathbf{y}$ may originate from different sensors and/or modalities, as reflected in the varying noise level $\sigma_k$. In practice, the weights are self-normalized to ensure $\sum_i w_i =1$ and computed in log-space to ensure numerical stability \citep{Aalstad2026}. Unlike sequential particle filters, which update only the current model state and parameters, the PBS applies the weights to the full temporal trajectory of each particle within the assimilation window, thereby propagating observational information both forward and backward in time \citep{Margulis2015AEstimation,Alonso-Gonzalez2022TheV1.0}.

All experiments use $1000$ ensemble members to ensure adequate prior sampling and minimise the risk of particle degeneracy, a condition in which weight collapses onto an implausibly small number of members, degrading the posterior representation \citep{Morzfeld2017}. Assimilation is performed independently at each observation time and for each glacier zone, and results are aggregated over the 12-year simulation period to obtain more robust inter-annual mean estimates.

\subsection{Experimental design}
\begin{figure}[!ht]
\begin{center}
\includegraphics[width=\linewidth]{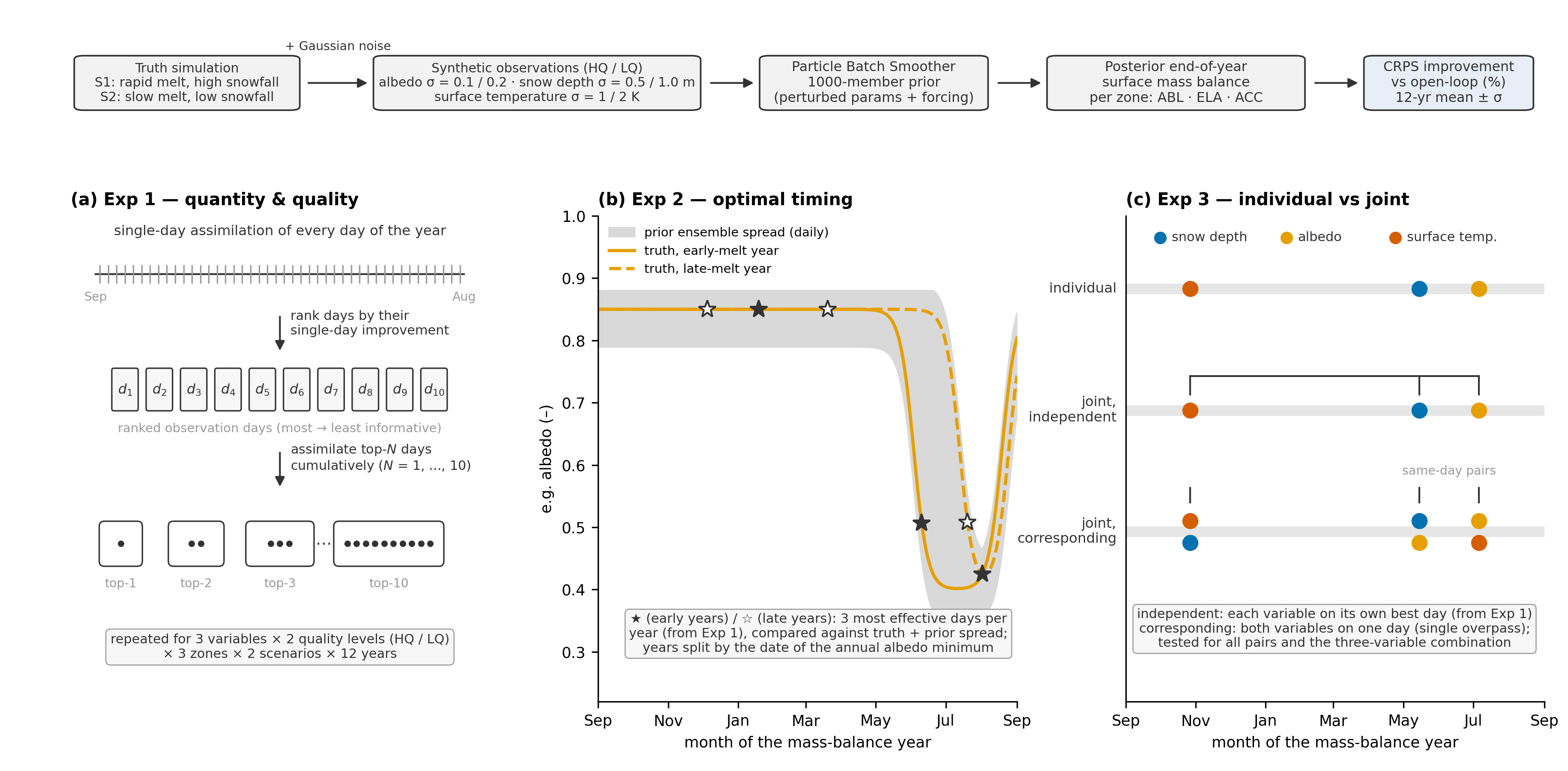}
\end{center}
\caption{Schematic overview of the experimental design. Top: the common synthetic twin pipeline shared by all experiments, in which synthetic observations of albedo, snow depth, and surface temperature are generated from a truth simulation under two climatic scenarios (S1, S2) and two quality levels (high and low observation noise), assimilated with the Particle Batch Smoother using a 1000-member prior ensemble, and evaluated as the percentage CRPS improvement in end-of-year surface mass balance relative to the open-loop prior, separately for the ablation (ABL), equilibrium line altitude (ELA), and accumulation (ACC) zones over 12 simulation years. \textbf{a)} Experiment 1 (quantity and quality):  the observation days are chosen by greedy forward selection, in which every remaining day of the year is assimilated in turn together with the already selected days and the day giving the largest further CRPS improvement is retained, iterated to $N=10$. This yields a nested sequence of optimized observation vectors of size $N=1,\dots,10$ and the corresponding improvement curve. The procedure is repeated for all combinations of observation type, quality level, glacier zone, and climatic scenario. \textbf{b)} Experiment 2 (optimal timing): the three jointly most effective observation days per year identified in Experiment~1 (stars) are analysed against the year-specific truth trajectory (illustrated here with albedo) and the daily prior ensemble spread (grey shading), with years partitioned into early and late melting regimes according to the date of the annual albedo minimum. \textbf{c)} Experiment 3 (individual vs joint): using each variable's single best observation day from Experiment~1, individual assimilation is compared with two joint strategies: the \textit{independent pair}, in which each variable is observed at its own optimal time, and the \textit{corresponding pair}, in which both variables are observed on the same day, as in a single satellite overpass. All pairwise combinations and the three-variable combination are tested. Marker colours denote the observation types, with blue for snow depth, yellow for albedo, and red for surface temperature.}
\label{fig:exp_design}
\end{figure}

Three sets of experiments are designed to address the study objectives (Figure~\ref{fig:exp_design}).

\textbf{Experiment 1: Observation quantity and quality.} For each observed variable (albedo, snow depth, surface temperature), we identify a highly informative set of observation days in each simulation year by greedy forward selection. First, a separate assimilation experiment is run for each day of the year, assimilating a single synthetic observation on that day, and the day giving the largest percentage improvement in CRPS relative to the prior is retained. This day is then held fixed, and a second observation is considered for the remaining days, now assimilating the retained first observation together with the candidate second one, ultimately retaining the second day whose addition gives the largest further CRPS improvement. Iterating up to ten observation days yields a nested sequence of observation vectors from size one to ten, along with the corresponding CRPS improvements as a function of observation count. Because each new observation day conditions on the already selected observation days, the $k$-th day is the best additional observation given the already selected $k-1$ days rather than the $k$-th most informative day considered in isolation. These two differ whenever observations are redundant, as is often the case for neighboring observations that are individually informative but largely redundant once one of them has been assimilated. Note also that the selected days are assimilated as a batch, so only the days contained in the observation vector, and not their ordering, affects the posterior. Such myopic greedy selection \citep{vanHove2026} is not guaranteed to recover the optimal set of ten days, which would require an intractable exhaustive search over all combinations but serves as a tractable approximation thereof. The procedure is repeated for high- and low-quality observations to examine the trade-off between observation precision and quantity, and results are aggregated over the 12 simulation years as mean improvement curves with interannual standard deviations.

\textbf{Experiment 2: Optimal observation timing.} To diagnose why certain combinations of observation days are most effective, the daily prior ensemble spread is computed for each variable throughout the simulation year. Years are partitioned into early and late melt seasons based on the calendar date of the annual albedo minimum. Years in which the minimum occurs before the multi-year median date are classified as early melting, and those after as late melting. The temporal distribution of the three most effective observation days per year from Experiment~1 is then plotted against the prior ensemble distribution and the truth trajectory, identifying whether high-impact observations coincide with periods of high model uncertainty. This analysis is conducted for the ablation zone, where the interplay between melt dynamics and ensemble divergence is most complex. Key differences for the ELA and accumulation zone are discussed in the context of the convergence results from Experiment~1.

\textbf{Experiment 3: Individual versus joint assimilation.} Using the single best observation day per year and variable from Experiment~1, individual and joint assimilation strategies are compared. Two joint strategies are tested. In the \textit{independent pair} strategy, the best observations for two different variables, each selected independently, are assimilated simultaneously, allowing each variable to be observed at its own optimal time. In the \textit{corresponding pair} strategy, the best observation from one variable is paired with the same-day observation of a second variable, representing a single coincident satellite overpass or ground visit. This design isolates whether temporal optimisation or observational diversity is the primary driver of joint assimilation benefit.

\subsection{Evaluation metrics}
Assimilation performance is quantified using the \textit{Continuous Ranked Probability Score} (CRPS). For a probabilistic forecast represented by a cumulative distribution function $P(x)$ and a scalar truth $y^\ast$, the CRPS is defined as \citep{Hersbach2000}
\begin{equation}
  \text{CRPS}(P, y^\ast) = \int_{-\infty}^{\infty}
  \bigl(P(x) - H(x-y^\ast)\bigr)^2\,\mathrm{d}x.
\label{eq:crps}
\end{equation}
where $H(\cdot)$ is the Heaviside function. The CRPS reduces to the mean absolute error when the forecast is deterministic and rewards both accuracy and sharpness of the probabilistic forecast. In this study, $P$ is approximated by the posterior ensemble of end of year surface mass balance using the Gaussian analytical equation from \citet{Gneiting2005} , and the percentage improvement relative to the prior CRPS is used as a measure of the added value of data assimilation. The CRPS strictly requires knowledge of the truth $y^\ast$, often substituted by a reference observation, and is therefore directly applicable in the synthetic twin experiment context of this study. 

\section{Results}

\subsection{Observation quantity and quality}

\begin{figure}[!ht]
\begin{center}
\includegraphics[width=\linewidth]{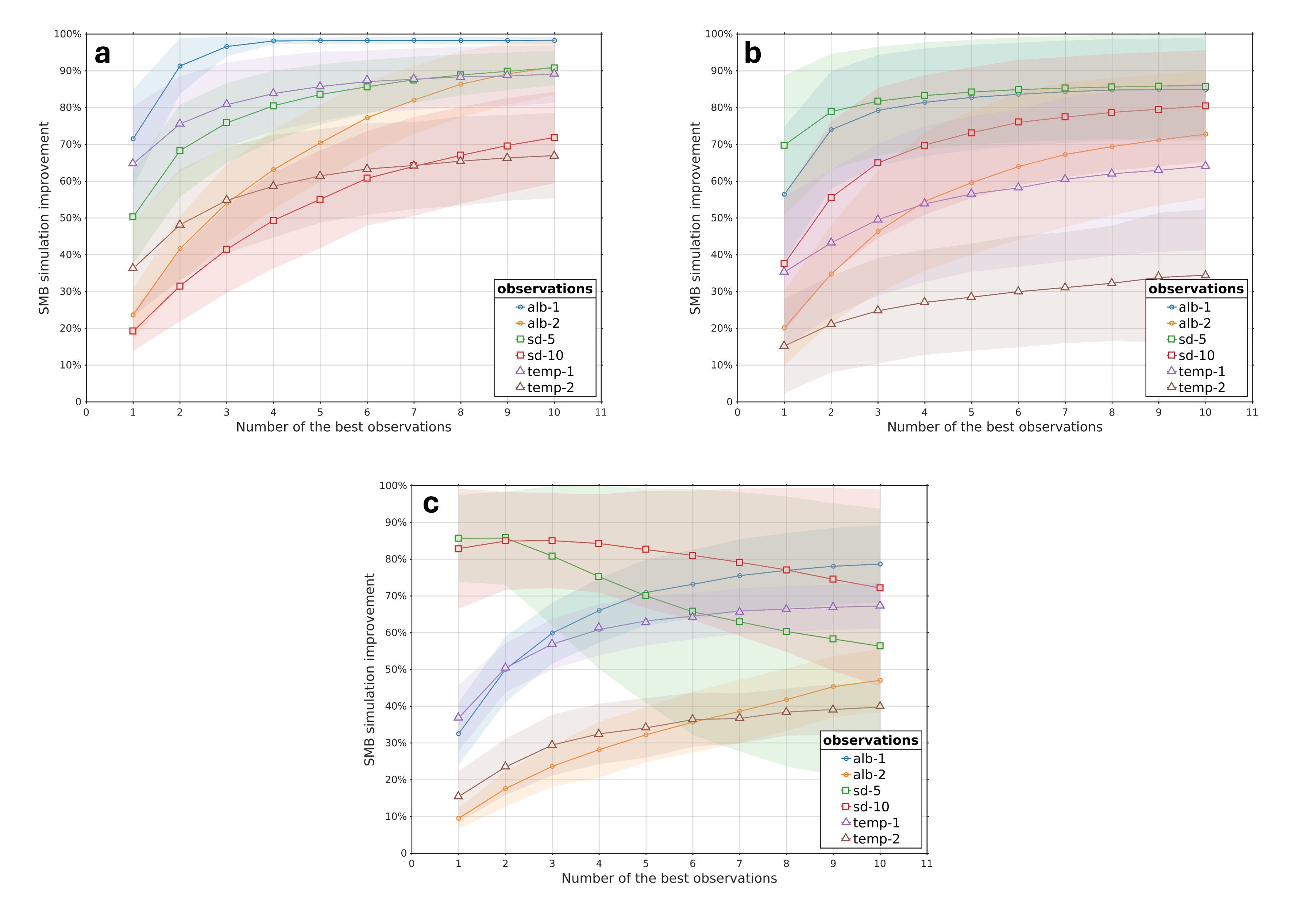}
\end{center}
\caption{Improvement in surface mass balance simulation, measured by percentage CRPS reduction relative to the open loop prior, as a function of the number of assimilated observations. Results are shaown for albedo, snow depth, and surface temperature in \textbf{a)} the ablation zone, \textbf{b)} the equilibrium line altitude zone, and \textbf{c)} the accumulation zone, under the climatic condition with rapid albedo evolution rate and high snowfall factor. Lines show the 12-year mean. Shaded bands show $\pm$1 standard deviation ($\sigma$) across years. Observation labels: high-quality albedo (alb-1), low-quality albedo (alb-2), high-quality snow depth (sd-5), low-quality snow depth (sd-10), high-quality surface temperature (temp-1), and low-quality surface temperature (temp-2).}
\label{fig:avg_syn1}
\end{figure}

The panels in figure \ref{fig:avg_syn1} show the improvements achieved using a varying number of optimized observation days across three distinct glacier zones. In the ablation zone, albedo is the most efficient single observation constraint. Assimilating just one high quality albedo observation per year yields $71\%$ improvement, converging to over $95\%$ with four observations. The efficiency advantage of high quality over low quality albedo is largest at low observation counts, as one low quality (higher noise) albedo observation achieves only $24\%$ improvement, but this gap closes progressively as the optimized observation count increases, reaching $90\%$ at ten low quality observations. This indicates that increasing observation quantity can partially compensate for lower observation quality, although with diminishing efficiency. Snow depth shows an intermediate pattern. A single high quality observation achieves $50\%$ improvement, reaching $90.8\%$ at ten observations, while low quality snow depth reaches parity with the single high quality result after four days and $70\%$ at ten. Surface temperature exhibits a similar quality-quantity tradeoff, with one high quality observation achieving $65\%$ improvement and ten observations approaching $90\%$. In contrast, ten low quality surface temperature observations are needed to match the performance of a single high quality one. Changes in the year-to-year standard deviation bands with increasing observation counts, particularly the widening observed for low-quality variables, reflect interannual variability in the information content of successively ranked observation days.

Near the ELA, the relative ranking of variables shifts so that snow depth becomes the most effective single observational constraint, yielding $70\%$ improvement from one high quality observation and converging at 85\% after five. This shift reflects the transitional hydrological dynamics near the ELA, where interannual variability in snow depth and the position of the ELA itself reflects whether a given year produces net accumulation or ablation. High quality albedo delivers $56\%$ improvement from one observation, converging to $80\%$ at ten. Low quality snow depth and albedo both reach around $70$--$80\%$ at ten observations, confirming a partially compensatory quantity effect of quantity over quality. Surface temperature shows the largest quality gap in this zone, as high quality observations provide roughly double the initial improvement of low quality observations, and neither fully converges at ten days.

The accumulation zone exhibits qualitatively distinct behaviour due to the presence of snow year-round. Snow depth delivers $>80\%$ improvement from a single observation regardless of quality, but performance degrades progressively when more than two observations are assimilated. This is a physically important finding. Because the snowpack never melts out, additional snow depth observations largely repeat the accumulation information that the first observation already provides, and assimilating more of them mainly concentrates the PBS weights onto very few ensemble members. The degradation is markedly faster for high quality snow depth, which falls from $86\%$ at one observation to $57\%$ at ten, than for low quality snow depth, which declines only from $83\%$ to $72\%$. This inverted quality ranking is a key signature of the underlying mechanism, since sharper observations concentrate the weights faster. Both curves nevertheless remain well above the prior at all observation counts, so the additional observations weaken the posterior relative to smaller observation sets rather than erasing the benefit of assimilation. The declining improvement is a manifestation of particle degeneracy, and the mechanism behind it is examined in the Discussion. Albedo and surface temperature do not show this degradation, in that both follow increasing convergence curves, with high quality data outperforming low quality data at all observation counts. High quality albedo achieves the highest ten-observation improvement among all variables in this zone, with low quality albedo initially lagging low quality surface temperature but overtaking it beyond six observations.

Overall, high quality observations provide more efficient single observation improvements than their low quality equivalents in all zones, although the advantage is marginal for snow depth in the accumulation zone. There, the degradation reverses the quality ranking beyond two observations, so that low quality snow depth ultimately outperforms high quality snow depth by a clear margin. Across all other cases, increasing observation quantity partially compensates for lower quality. This compensation is clear but imperfect. Achieving $>90\%$ improvement with low quality observations requires two to five times as many observations as with high quality observations.

\begin{figure}[!ht]
\begin{center}
\includegraphics[width=\linewidth]{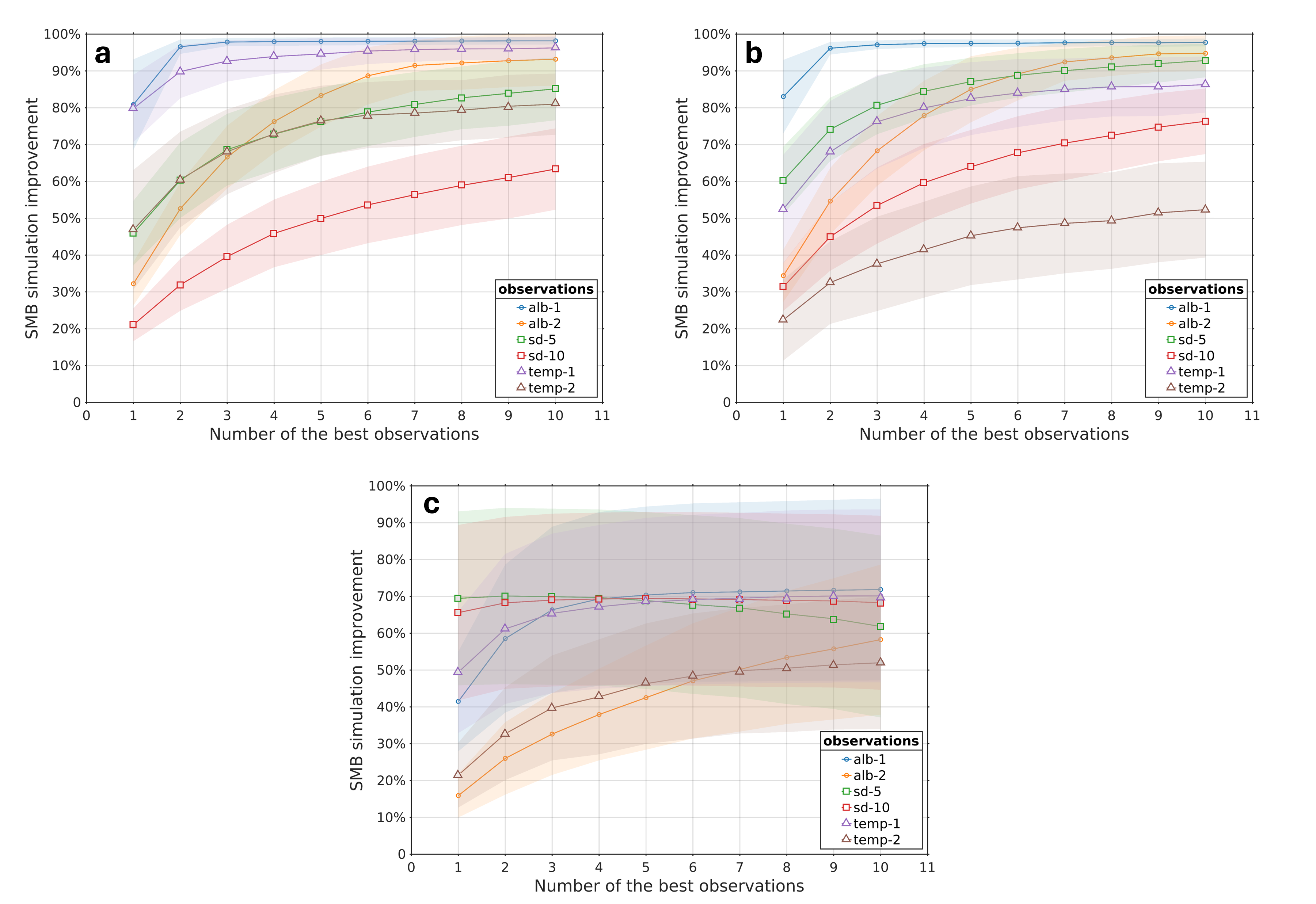}
\end{center}
\caption{As Figure \ref{fig:avg_syn1}, but under the climatic condition with slow albedo evolution rate and low snowfall factor (S2).}
\label{fig:avg_syn2}
\end{figure}

Figure \ref{fig:avg_syn2} shows the same convergence analysis under the slow albedo evolution and low snowfall scenario.

The S2 scenario produces a more differentiated shift in assimilation performance. High quality albedo and surface temperature observations become more effective at the single observation level compared to S1, while snow depth observation effectiveness weakens across all zones.

In the ablation zone (Fig. \ref{fig:avg_syn2}a), high quality albedo (alb-1) and high quality surface temperature (temp-1) are essentially tied as the most effective observed variables, yielding $81\%$ and $80\%$ improvement, respectively, from a single best observation, both higher than their S1 counterparts of $71\%$ and $65\%$. High quality snow depth (sd-5), by contrast, drops to $46\%$ from a single observation compared to ${\approx}50\%$ in S1, becoming clearly the least effective high quality variable in this zone. The dominant feature of S2 in the ablation zone is therefore a large gap between the two leading variables (alb-1 and temp-1) and snow depth, a contrast absent in S1 where all three high quality variables were performing more closely. At ten observations, alb-1 and temp-1 converge to $90$--$93\%$, while sd-5 reaches $80$--$85\%$. The year to year standard deviation bands are narrower than in S1, reflecting reduced interannual variability under the slow-evolution scenario.

In the ELA zone (Fig. \ref{fig:avg_syn2}b), albedo emerges as the dominant constraint by a clear margin, achieving $83\%$ improvement from a single high quality observation, a dramatic increase from its S1 value of ${\approx}56\%$. Snow depth drops from its S1-leading position of ${\approx}70\%$ to $60\%$, while surface temperature falls to  $52\%$. The ranking in S2 at the ELA is therefore alb-1 $>$ sd-5 $>$ temp-1, in contrast to the S1 ELA ordering (sd-5 $>$ alb-1 $>$ temp-1). The sharp rise of albedo effectiveness reflects the prolonged period of ensemble divergence across the slow melt season transition, which makes the single best albedo observation in a slow-evolution year exceptionally discriminating. The decline of snow depth performance results from reduced interannual snowpack variability and magnitude under this low snowfall scenario, which limits the ability of noisy snow depth observations to distinguish between ensemble members near the equilibrium line. At ten observations, alb-1 converges to approximately $90$--$93\%$, sd-5 to  $83$--$87\%$, and temp-1 to $78$--$82\%$.

In the accumulation zone (Fig. \ref{fig:avg_syn2}c), the snow depth degradation pattern persists but proceeds considerably more gradually than in S1. The performance of high quality snow depth constraints starts at $70\%$ from a single observation and declines slowly, reaching approximately $69\%$ by five observations and $62\%$ by ten. This is a modest and steady decline rather than the steep collapse seen in S1, confirming that the degradation is a structural feature of PBS particle degeneracy in snow-retaining zones but that its severity depends on the climatic scenario. Low quality snow depth, by contrast, holds nearly constant between $65$ and $70\%$ across all observation counts, again consistent with slower weight concentration under larger observation noise. Albedo and surface temperature maintain monotonically increasing convergence curves. High quality albedo reaches $72\%$ at ten observations and high quality surface temperature $70\%$.


\subsection{Timing}
\begin{figure}[!ht]
\begin{center}
\includegraphics[width=\linewidth]{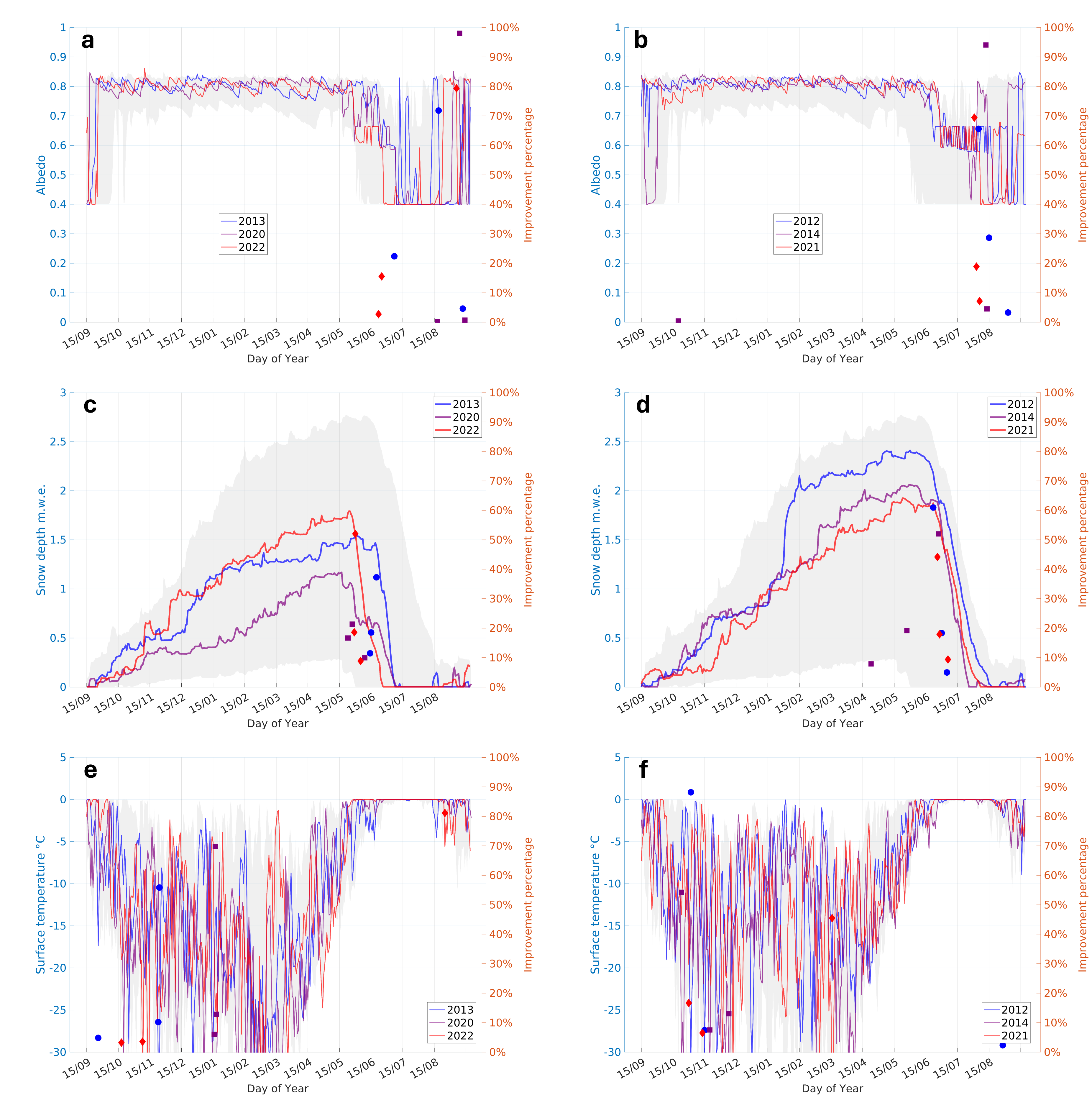}
\end{center}
\caption{Impact of individual variable assimilation on glacier surface mass balance simulation accuracy in the ablation zone under the rapid albedo evolution and high snowfall scenario. Panels are organised by melting regime and assimilated variable: early melting years (left column: a, c, e) and late melting years (right column: b, d, f) for (a--b) albedo, (c--d) snow depth [m w.e.], and (e--f) surface temperature [$^\circ$C]. Left $y$-axes show the physical values of the assimilated observations. Grey shaded regions denote the prior ensemble spread. Solid coloured lines represent the truth observations for each year. Right $y$-axes show the percentage CRPS improvement [$\%$] following assimilation. Scattered markers, colour-coded by year, indicate the improvement from assimilating a specific observation on that day. For visual clarity, only the three most impactful observations are shown per year.}
\label{fig:truth_contri_syn1_ABL_high}
\end{figure}

Figure \ref{fig:truth_contri_syn1_ABL_high} shows the temporal distribution and improvement contribution of albedo (a, b), snow depth (c, d), and surface temperature (e, f) in the ablation zone under the rapid albedo evolution and high snowfall scenario, partitioned into early and late melting years. For each year, only the three most impactful observations are plotted. The grey shading represents the prior ensemble spread (standard deviation across the $1000$ members ensemble) throughout the year, and the coloured lines represent the truth trajectories for each individual year. Wide grey shading indicates high prior uncertainty, where an observation can strongly discriminate between ensemble members and help achieve large surface mass balance CRPS improvements. Narrow shading indicates low prior uncertainty, where observations have limited scope for further constraining the ensemble. Thereby, the temporal structure of prior ensemble spread helps govern when observations are most effective and is directly linked to the information gain achievable by data assimilation.

\textit{Albedo.} The timing of the most effective albedo observations cluster tightly around the sharp snow-to-ice transition in the truth trajectory, the period when the truth albedo drops rapidly from high snow-covered albedo to lower bare-ice albedo values. In early melting years, this transition occurs earlier in summer and so the best observations concentrate between mid-June and mid-July during the  albedo decline, with the highest single-observation improvement reaching nearly $98\%$. Secondary contributions appear in mid-August to mid-September during subsequent albedo fluctuations from intermittent summer snowfall. In late melting years, the sharp transition is delayed by four to six weeks and the optimal window shifts to late July through mid-August, with peak improvements up to $90\%$. This timing shift follows the year-specific truth trajectory. The prior parameter distribution is identical in every year, while each year's prior ensemble and truth are driven by that year's meteorological forcing, which sets the date of the transition from snow to bare ice. Observations acquired while the truth crosses this transition are the most informative because ensemble members with different combinations of the parameter set expose bare ice at different dates. The year-specific date of the sharp albedo decline in the truth therefore anchors the optimal observation window. Outside this transition period, when the truth albedo is either uniformly high in winter or has already stabilised at bare-ice values, information gain drops sharply, as the truth no longer occupies a discriminating position relative to the ensemble. This 4-6 week interannual shift confirms that fixed calendar observation strategies systematically miss the most informative period in at least one melting regime.

\textit{Snow depth.} The optimal snow depth observations are anchored to the period of rapid snowmelt in the truth trajectory, when observed snow depth is falling steeply and the truth value most clearly distinguishes between ensemble members with different snowpack evolution paths. In early melting years, this period concentrates between mid-May and mid-June, when rapid melt translates small inter-ensemble member differences in snow depth into large divergences in the timing of the  melt-albedo feedback and the subsequent emergence of bare-ice. In late melt years, the optimal window shifts to mid-June through mid-July, tracking the delayed onset of rapid melt in the truth. An exception in 2014 places two high-impact observations during the accumulation period, suggesting that years with anomalous snow accumulation events can produce a secondary sharp transition in truth snow depth early in the season. The peak single observation improvement from snow depth is lower than for albedo, ranging from $20\%$ to $60\%$, reflecting the weaker but seasonally specific coupling of snow depth to end-of-year mass balance in the ablation zone.

\textit{Surface temperature.} Surface temperature presents a distinct case. During the melt season, the surface of every ensemble member is pinned at the melting point, so the prior temperature spread collapses to nearly zero and summer observations carry almost no information. The informative period is therefore confined to the cold part of the year, when ensemble members cool and refreeze at different rates and the temperature spread is wide. The best single observations exceed $80\%$ improvement and fall mostly between mid-September and mid-January, with late melting years concentrating in mid-October to mid-November. Spread alone, however, cannot explain this timing, because the spread remains wide throughout the accumulation season while the most effective days cluster tightly. That clustering follows the truth trajectory. The sharp cooling transition at the onset of the accumulation season, together with occasional pronounced temperature anomalies, places the truth in a discriminating position within the widened ensemble, and the truth temperature at these moments most clearly separates the member trajectories. In one early melting year, a cold anomaly from fresh snowfall in mid-August already provides this discrimination near the end of the ablation season. This cold-season truth transition encodes year-specific information about the snowpack thermal state that propagates into the subsequent melt season.

\begin{figure}[!ht]
\begin{center}
\includegraphics[width=\linewidth]{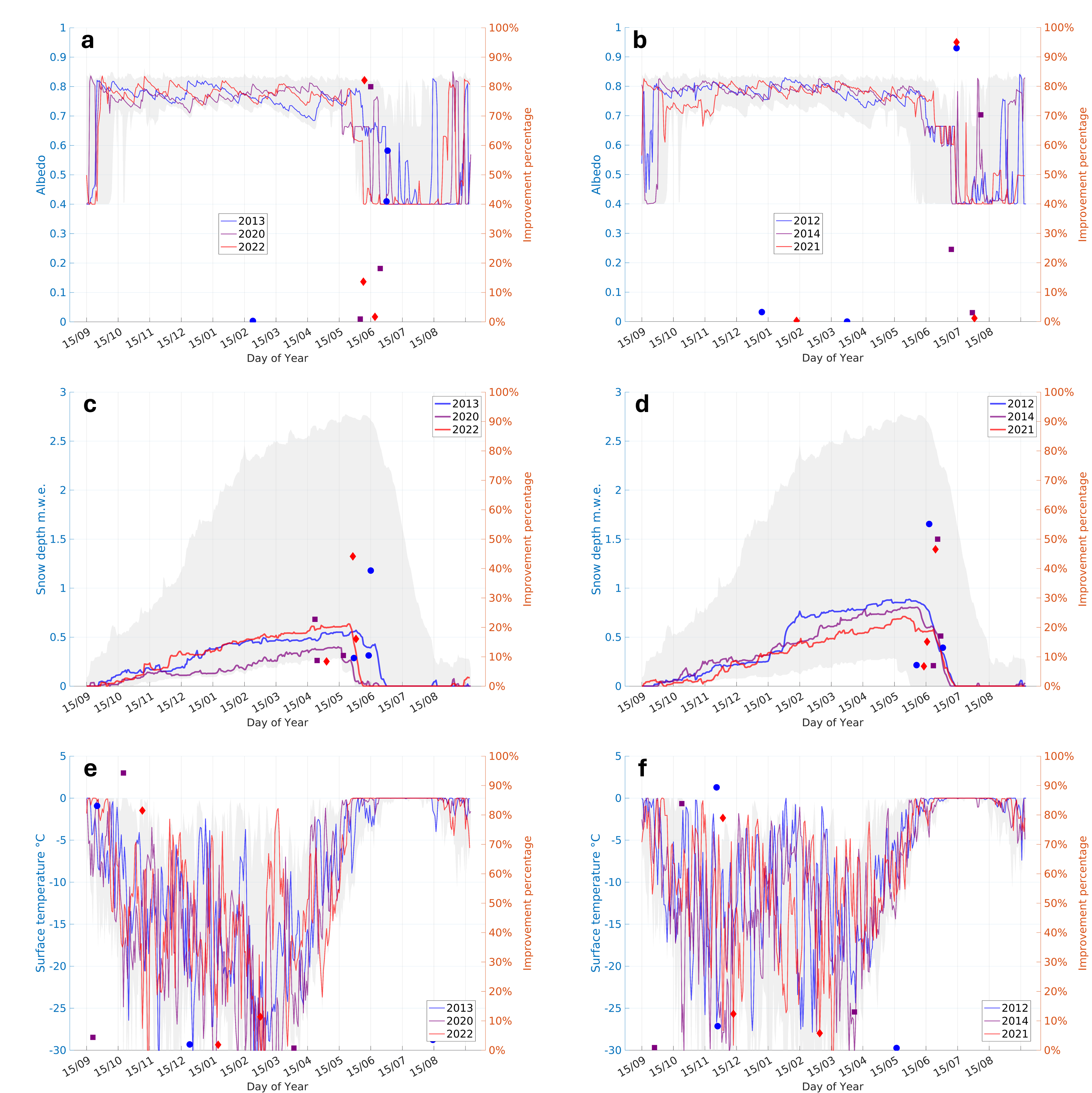}
\end{center}
\caption{Impact of individual variable assimilation on glacier surface mass balance simulation accuracy in the ablation zone under the slow albedo evolution and low snowfall scenario. Layout is identical to Figure~\ref{fig:truth_contri_syn1_ABL_high}: panels are organised by melting regime (left column: early melting years a, c, e and right column: late melting years b, d, f) and by assimilated variable ((a--b) albedo, (c--d) snow depth [m w.e.], (e--f) surface temperature [$^\circ$C]). Grey shaded regions show the prior ensemble spread. Coloured lines show the truth trajectories. Scattered markers show the CRPS improvement from assimilating the three most impactful observations per year.}
\label{fig:truth_contri_syn2_ABL_high}
\end{figure}

Figure~\ref{fig:truth_contri_syn2_ABL_high} presents the same timing analysis for the ablation zone under the slow albedo evolution and low snowfall scenario. This scenario provides a controlled test of the truth-trajectory argument. The prior ensemble is identical in the two scenarios because it is generated from the same large $1000$ member parameter prior ensemble and the same meteorological forcing. Any difference in optimal observation timing between S1 and S2 must therefore originate from the truth trajectories.

The timing of the most effective observations remains anchored to the seasonal transitions of the truth under this scenario. For albedo, the low snowfall factor produces a thin snowpack that melts out abruptly, so the transition from snow to bare ice remains sharp even though the albedo itself evolves slowly. The most effective albedo observations concentrate at this transition, which occurs in early to mid June in early melting years and around mid July in late melting years, with peak improvements of $82\%$ and $95\%$, respectively. Before the transition, the truth albedo is uniformly high. Thereafter, the truth sits at bare ice values, and observations outside the transition window contribute little. For snow depth, the most effective observations remain within the depletion phase, between mid April and early June in early melting years and in the second half of June in late melting years. Peak improvements of $39$ to $55\%$ are slightly lower than under S1, consistent with the thinner truth snowpack providing a weaker discriminating signal. For surface temperature, the accumulation season preference is preserved. The best observations fall between late September and early December and reach improvements of $80$ to $94\%$, confirming that the cooling transition at the onset of the accumulation season is a structurally similar feature in both scenarios. Across all three variables, optimal timing follows the year-specific transitions of the truth in both scenarios, and the interannual shift between early and late melting years persists under S2. Since the prior ensemble is shared between the scenarios, these timing differences cannot arise from the prior and must be inherited from the truth trajectories encoded in the noisy observations.


\subsection{Individual and Joint assimilation}
\begin{figure}[!htbp]
\begin{center}
\includegraphics[width=\linewidth]{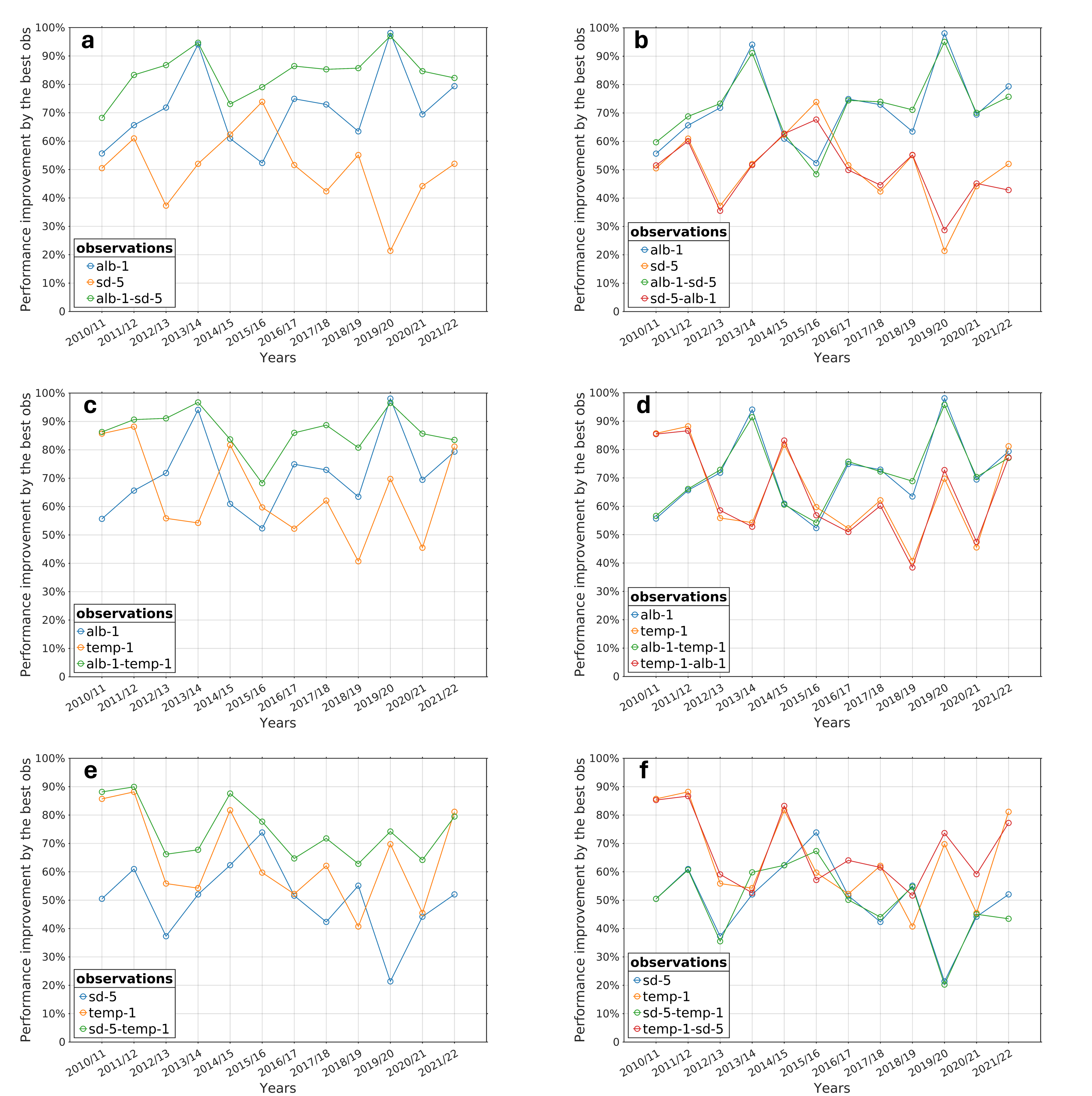}
\end{center}
\caption{Year-by-year comparison of simulated surface mass balance CRPS improvement by individual and joint assimilation strategies in the ablation zone under the rapid albedo evolution and high snowfall scenario, across all 12 simulation years. Left panels show the \textit{independent pair} strategy, where each variable's best observation is selected at its own optimal day and both are assimilated jointly. Panel (a) shows albedo (alb-1), snow depth (sd-5), and their joint result (alb-1-sd-5). Panel (c) shows albedo (alb-1), surface temperature (temp-1), and their joint result (alb-1-temp-1). Panel (e) shows snow depth (sd-5), surface temperature (temp-1), and their joint result (sd-5-temp-1). Right panels show the \textit{corresponding pair} strategy, where the best observation of one variable is paired with the same-day observation of the other, and vice versa. Panel (b) shows alb-1, sd-5, the pair alb-1-sd-5 with the best albedo day and co-located snow depth, and the pair sd-5-alb-1 with the best snow depth day and co-located albedo. Panel (d) shows alb-1, temp-1, alb-1-temp-1, and temp-1-alb-1. Panel (f) shows sd-5, temp-1, sd-5-temp-1, and temp-1-sd-5.}
\label{fig:duo_syn1_ABL}
\end{figure}

Figure~\ref{fig:duo_syn1_ABL} compares individual and joint assimilation performance for all three pairwise variable combinations in the ablation zone under the rapid albedo evolution scenario, shown year by year across the 12-year period. The left panels (a, c, e) show the independent pair strategy, in which the best observation for each variable is selected at its own optimal time and then assimilated jointly. The right panels (b, d, f) show the corresponding pair strategy, in which two variables are observed on the same day. All panels include the individual performance of each variable for direct comparison.

\textit{Albedo and snow depth (Fig. \ref{fig:duo_syn1_ABL}  a--b).} The best annual albedo observation (alb-1) delivers improvements of approximately $50$--$90\%$ across the 12 years, consistently outperforming the best snow depth observation (sd-5, approximately $20$--$70\%$) in all but one year. The independent (i.e., asynchronous) joint assimilation (alb-1-sd-5, Fig.\ref{fig:duo_syn1_ABL}  a) lies at or above both individual lines in every year without exception, reaching above $70\%$ improvement in all years. The physical basis for this gain is that the optimal albedo observation from the melt season transition constrains summer surface energy balance, while the optimal snow depth observation from the rapid snowmelt phase constrains runoff timing and initial melt conditions. These two windows are temporally distinct, so the information they carry is largely non-redundant. The sole year in which the joint result matches rather than exceeds individual albedo is the year in which alb-1 alone achieves its highest improvement (${\approx}90\%$), where albedo already nearly fully constrains the prior. The corresponding pair strategy (Fig. \ref{fig:duo_syn1_ABL}b) tells a starkly different story. Assimilating the best albedo with the same-day snow depth (alb-1-sd-5, green) produces a result that closely shadows alb-1 alone without systematic gain, while assimilating the best snow depth with the same-day albedo (sd-5-alb-1, red) closely shadows sd-5 alone. Each corresponding pair is dominated entirely by its leading variable, confirming that same-day observations provide correlated rather than complementary information.

\textit{Albedo and surface temperature (Fig. \ref{fig:duo_syn1_ABL} c--d).} Surface temperature (temp-1) shows substantially higher interannual variability than snow depth, ranging from approximately $40\%$ to $90\%$ improvement depending on the year, compared to the more stable $50$--$90\%$ of albedo. In years where temp-1 is high, the independent joint assimilation (alb-1-temp-1, Fig. \ref{fig:duo_syn1_ABL} c) lies well above alb-1 alone, producing the largest year-specific gains of any pairwise combination. In years where temp-1 is lower, the joint result remains at or above alb-1, confirming that the independent pair strategy never degrades performance. This guarantee holds because when the additional variable carries little information in a given year, the PBS effectively downweights it, preserving the performance of the leading variable. The corresponding pair strategy (Fig. \ref{fig:duo_syn1_ABL} d) removes the cold season advantage of surface temperature entirely. Assimilating temperature on the same day as the best albedo observation yields a result that shadows alb-1 closely (alb-1-temp-1, green), while the reverse pair (temp-1-alb-1, red) shadows temp-1. Neither corresponding pair achieves systematic improvement over its individual components.

\textit{Snow depth and surface temperature (Fig. \ref{fig:duo_syn1_ABL} e--f).} Snow depth (sd-5) and surface temperature (temp-1) have the most temporally distinct optimal observation windows of any pair, with rapid spring snowmelt for snow depth and the accumulation season for surface temperature. This makes them the combination with the greatest potential temporal complementarity. The independent joint assimilation (sd-5-temp-1, Fig. \ref{fig:duo_syn1_ABL} e) is consistently at or above both individual lines. In years where temp-1 is high, the joint substantially exceeds sd-5 alone, while in years where temp-1 is low, the joint follows the higher of the two individual results. The corresponding pair strategy (Fig. \ref{fig:duo_syn1_ABL} f) again fails to capture this complementarity. Assimilating the best snow depth day together with same-day temperature (sd-5-temp-1, green) closely tracks sd-5 alone, while the reverse pair (temp-1-sd-5, red) closely tracks temp-1. The large temporal mismatch between the optimal timing of these two observed variables means that constraining one to the other's best day sacrifices almost all of its individual information gain.

\begin{figure}[!htbp]
\begin{center}
\includegraphics[width=\linewidth]{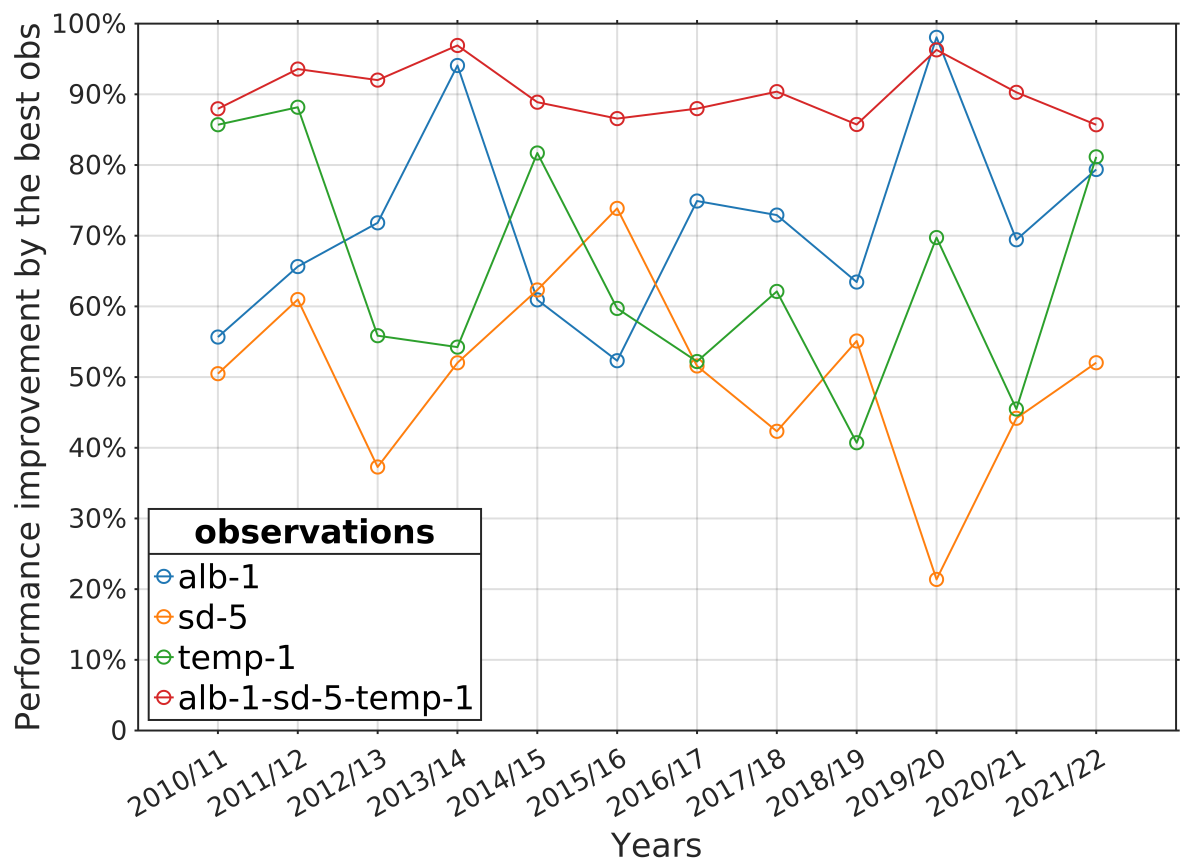}
\end{center}
\caption{Year-by-year comparison of surface mass balance simulation improvement by individual and three-variable joint assimilation in the ablation zone under the rapid albedo evolution and high snowfall scenario, across all 12 simulation years. Each variable's single best observation is selected at its own optimal day: albedo (alb-1), snow depth (sd-5), and surface temperature (temp-1). The three-variable independent joint assimilation (alb-1-sd-5-temp-1) assimilates all three best observations simultaneously, each at its own optimal time. The joint result lies at or above the strongest individual variable in every year and remains within a narrow high band across the full period.}
\label{fig:trio_syn1_ABL}
\end{figure}

\textit{Three-variable joint assimilation.} Figure~\ref{fig:trio_syn1_ABL} shows the three-variable independent joint assimilation, in which the independently selected best observations for albedo, snow depth, and surface temperature are assimilated simultaneously, each at its own optimal time. This strategy represents the upper bound achievable with a single observation per variable type. The joint result (alb-1-sd-5-temp-1) lies at or above the strongest individual variable in every year and remains within a narrow high band of approximately $85$ to $97\%$ improvement across all 12 years. In the single year where albedo alone nearly saturates the improvement (2019/20, approximately $98\%$), the joint result coincides with it, and in all other years it sits at or above the leading individual variable. This stability across years is the defining feature of the three-variable combination. When an individual variable performs poorly in a given year, the joint result is sustained by the remaining two. The clearest example is 2019/20, where the best snow depth observation collapses to approximately $21\%$, yet the joint result reaches approximately $97\%$, anchored by the strong albedo constraint that year. Each of the three observations constrains a temporally distinct process. Albedo constrains the surface energy balance during the melt season, snow depth constrains the spring snowmelt dynamics, and surface temperature constrains the initialisation of the winter energy and mass budget. Together, they provide  more complete constraints on the dominant processes governing annual surface mass balance in the ablation zone. This result demonstrates that the full benefit of multivariate joint data assimilation can be better realised when each variable is observed at its own independently optimised time.

\section{Discussion}

\subsection{Quantity versus quality: the compensation mechanism and its limits}

The consistent finding that an increased quantity of observations can compensate for reduced observation quality is explained by the cumulative information content of multiple independent observations. Each additional observation, even if noisy, contributes incremental information via the likelihood to the PBS weight update, progressively narrowing the posterior even when no single observation is highly discriminating \citep{Margulis2015AEstimation,Robinson2018ImprovingObservations}. The compensation is therefore most effective when successive observations provide independent information, which requires that they fall on different days, ideally spanning different phases of the mass balance cycle.

In the ablation zone, the performance gap between high quality and low quality albedo is most pronounced at low observation counts (one to three), where noise prevents the PBS from strongly differentiating ensemble members. As more observations are added, the cumulative signal to noise ratio grows and the gap narrows. A similar pattern holds for snow depth and surface temperature, though with variable rates of compensation. This behaviour is consistent with previous applications of particle methods to snow and glacier systems, where the marginal gain from each additional observation diminishes once the leading constraint has captured the dominant source of ensemble spread \citep{Smyth2019ParticleSWE,Guidicelli2024}.

In the accumulation zone, the anomalous degradation of snow depth assimilation with increasing observation count signals a breakdown of the compensation mechanism. Because the snowpack persists throughout the year, snow depth observations from different parts of the season act as redundant constraints on the same accumulation trajectory rather than as independent pieces of information. Each added observation therefore contributes little new information about the end of year mass balance, yet it still sharpens the joint likelihood, concentrating weight on the few ensemble members that best fit the observed depth sequence. A close fit to the observed snow depth does not imply a close fit to the mass balance, for two reasons. First, the fit is to the noisy observations rather than to the true snowpack, so the favoured members carry a small accumulation bias that integrates into the annual mass balance. Second, snow depth in this zone constrains the snowfall factor but carries little information about the albedo evolution rate, to which the accumulation zone mass balance remains sensitive through summer melt. The strong improvements achieved by albedo observations in this zone confirm this sensitivity. The weight collapse therefore pins down the well observed parameter dimension while leaving the poorly observed dimension to the few surviving members, whose albedo parameter values are effectively arbitrary. Even when the posterior mean remains close to the truth, the collapsed spread is itself penalised by the CRPS, which rewards calibrated uncertainty. Consistent with this interpretation, the degradation is consistently stronger for the sharper high quality observations, whose likelihood concentrates the weights more rapidly, and in the low snowfall scenario the noisier low quality observations barely degrade at all. The same snow depth observations instead improve performance with increasing count in the ablation and equilibrium line zones, where successive observations sample distinct phases of the seasonal cycle and are therefore not redundant. Particle degeneracy, the collapse of effective ensemble size due to extreme weight concentration, is a well-known failure mode of particle methods and has been identified in glacier and snowpack assimilation contexts when the observation operator is highly sensitive to the assimilated variable \citep{Landmann2021AssimilatingFilter, Aalstad2026}. In the present case, this failure mode is triggered not by a single highly constraining observation, but by the cumulative effect of multiple redundant snow depth observations whose joint likelihood sharpens faster than the information they add. Diagnosing this condition in real-world settings, where the truth is unknown, requires monitoring the effective sample size or the concentration of posterior weights after each assimilation step. A rapid collapse of weight onto a small number of ensemble members should be treated as a warning of potential degeneracy rather than a sign of successful constraint \citep{Morzfeld2017}.

\subsection{Alignment of truth value and optimal observation timing}

The analysis leads to a clear conclusion about what governs optimal observation timing. It is the structure of the truth trajectory, specifically the timing and sharpness of its seasonal transitions, that determines when a single observation achieves maximum information gain. Prior ensemble spread provides a necessary background condition, in that the ensemble must be spread enough for an observation to meaningfully differentiate among members. It does not, however, identify which moment within a broad season of elevated uncertainty is most valuable. That specificity comes from the underlying truth signal that is encoded in noisy observations.

The strongest evidence for this claim is the comparison between early and late melting years. The prior parameter distribution is the same in every year, so the four to six week shift in optimal albedo timing between the two melting regimes cannot reflect any change in the assumed parameter uncertainty. The shift is instead imposed by the year-specific meteorological forcing through the truth trajectory it generates. The best observation is the one that captures the truth at its most rapidly changing and most discriminating moment, and that moment varies with the individual year's melt dynamics \citep{Davaze2018MonitoringData, Dumont2012LinkingData, Ye2024UnveilingGlaciers}.

The S2 scenario provides a complementary and controlled test, because the prior ensemble is identical in the two scenarios. Any change in optimal observation timing between the scenarios must therefore be inherited from the truth trajectories, and the test confirms their anchoring role. In the ablation zone the thin snowpack under low snowfall melts out abruptly, the truth albedo transition remains sharp despite the slow albedo evolution, and the most effective albedo observations stay concentrated at the year-specific transition dates. For snow depth, the depletion phase in the truth continues to anchor the optimal window, while the thinner snowpack weakens the discriminating signal and lowers the achievable improvements \citep{VanPelt2019A1957-2018, Pramanik2019ComparisonPrecipitation}. For surface temperature, the accumulation-season cooling transition in the truth identifies the optimal window in both scenarios. Surface temperature illustrates both roles at once. During the melt season, the surface of every member sits at the melting point, the spread vanishes, and no temperature observation can discriminate, which confines the informative period to the cold season. Within that season the spread stays wide for months, so it cannot single out the best days, and it is the cooling transition of the truth that does \citep{stby2014SevereSvalbard, Karner2013ASvalbard}.

This framing has a direct practical implication. Adaptive observation scheduling should be driven by monitoring the real-time trajectory of each observed variable relative to its historical ensemble distribution. Therein, the goal would be to detect the onset of sharp transitions (the beginning of rapid albedo decline, the onset of snowmelt, the autumn temperature inflection) and concentrate observations at those moments rather than spread them uniformly across a broad seasonal window. Fixed calendar strategies fail precisely because these transitions vary interannually by several weeks. Spread-guided scheduling that also tracks truth-trajectory sharpness would consistently capture the most informative period across all years \citep{Carrassi2018DataPerspectives, Alonso-Gonzalez2023ExploringExperiment}. Future work could formalise this through information-theoretic measures \citep{Teweldebrhan2019,vanHove2026}, such as the information gain measured as the reduction in Shannon entropy of the posterior weight distribution compared to that of the prior \citep{Lindley1956}, which would provide operationally accessible diagnostics for identifying the onset of optimal observation windows.

\subsection{Joint versus individual assimilation: temporal diversity over observational diversity}

The central insight from the joint assimilation experiments is that the benefit of multi-variable assimilation derives from the temporal diversity of the observation types, not from the addition of more variables. This is most evident in the three-variable independent joint assimilation (Figure~\ref{fig:trio_syn1_ABL}), which stays within a narrow high band of improvement across all 12 years even when a single variable performs poorly in a given year, because the remaining two variables sustain the constraint. When each variable is observed at its own optimal time, exploiting the seasonally distinct optimal windows of albedo, snow depth, and surface temperature, the resulting joint posterior integrates near-independent constraints on the three dominant components of annual glacier surface mass balance. The value of combining multiple observation types in glacier and snowpack data assimilation has been demonstrated in several recent studies \citep{Montzka2012MultivariateReview, Largeron2020TowardReview, Alonso-Gonzalez2022TheV1.0, Navari2021Reanalysis20002014,Mazzolini2024Spatio-temporalAltimeter}, but the present results show that realising this value is facilitated by the assimilation of each variable being scheduled at its own seasonally optimal time rather than co-located in time with other variables. When variables are instead constrained to be observed simultaneously, the joint assimilation is dominated by whichever variable has higher individual utility on that shared day, and the complementary information of the other variable is largely lost.

The contrast between the independent and corresponding pair strategies isolates this mechanism most directly. Under the independent strategy, where each variable is observed at its own optimal day, the joint result lies at or above both individual results in every year and never degrades the performance of the stronger single variable. This guarantee follows from the weight update of the PBS. When the second variable carries little additional information in a given year, its likelihood is close to uniform across the ensemble, the posterior weights remain governed by the more informative variable, and the joint result cannot fall below the better individual constraint \citep{Fowler2012MeasuresAssimilation}. Under the corresponding strategy, where the two variables are observed on the same day, this property is lost. Each corresponding pair shadows whichever variable dominates on the shared day, because an observation taken outside its own optimal window samples a dynamical state that the leading variable already constrains and therefore carries correlated rather than complementary information \citep{Fowler2012MeasuresAssimilation}. The gain from adding a second variable thus depends almost entirely on whether it is observed within its own seasonally distinct window, not on the simple fact of adding it. This is consistent with the strong sensitivity of snow assimilation skill to observation timing reported for snow depth \citep{Smyth2020ImprovingTiming, Smyth2019ParticleSWE,Guidicelli2024}. The three-variable independent joint assimilation extends this logic to its upper bound. Because the three optimal windows are mutually distinct, the combined constraint is close to additive, and the year-to-year stability of the joint result in Figure~\ref{fig:trio_syn1_ABL} reflects the low redundancy among the three temporally separated observations.

This result has a clear practical implication for satellite-based monitoring. The commonly used assumption that simultaneous multi-variable retrieval from a single overpass provides the best joint assimilation input is incorrect. A single satellite overpass provides high observational diversity but zero temporal diversity, and our results show that temporal diversity can be the more valuable property. Satellite-based monitoring in high-latitude environments is further constrained by cloud contamination and revisit-time limitations \citep{Aalstad2020,Kotarba2022ImpactMissions}, which reduce the effective frequency of cloud-free acquisitions and underscore the need to prioritise observations that fall within the seasonally optimal windows of each variable. The timing of albedo observations relative to the melt transition is particularly critical. Variational and ensemble-based assimilation studies have shown that albedo observations during the melt season, when surface reflectance is transitioning from snow to ice, carry substantially more information for constraining mass balance than observations acquired outside this window \citep{Dumont2012VariationalGlacier, Bertoncini2024AssimilationHydrology}. Designing monitoring systems that schedule different sensor types or revisit times to target the seasonally distinct optimal windows for each variable, rather than requiring simultaneous coverage, would more efficiently exploit the information content available from remote sensing \citep{Alonso-Gonzalez2022TheV1.0, Alonso-Gonzalez2023ExploringExperiment, Berthier2023MeasuringReview}.

\subsection{Limitations and outlook}

Several limitations of the present study should be acknowledged. First, this work is based entirely on synthetic twin experiments, in which the truth is generated by the same model used to construct the prior ensemble. This is a standard framework in observing system simulation experiments (OSSEs) \citep{Masutani2010,Alonso-Gonzalez2023ExploringExperiment} that inherently excludes structural model error. Consequently, differences between how the model represents physical processes and how those processes occur in nature are absent from the experimental design. In real applications, the model-truth discrepancy may be larger and more complex than the noise levels considered here, potentially reducing the single observation improvements and altering the optimal timing windows identified. Real satellite observations carry spatially correlated retrieval errors, systematic biases, and cloud-induced data gaps \citep{Berthier2023MeasuringReview, Paul2017ErrorProject} that are not captured by the independent Gaussian noise model used here. Extending the framework to real satellite-derived observations is therefore an essential and challenging next step.

Second, the analysis is conducted on a single glacier in Svalbard with a specific geometry and climatic setting. Mass balance dynamics and the relative importance of different observation types vary substantially across Svalbard's glacier population, which spans a wide range of sizes, elevations, aspects, and thermal regimes \citep{Schuler2020ReconcilingBalance, Moller2018Differing19002010}. The relative importance of albedo, snow depth, and surface temperature as observational constraints, and the seasonal timing of their optimal windows, will further vary with glacier hypsometry, debris cover, and polythermal behaviour across the broader Arctic \citep{VanPelt2021AcceleratingEnsemble, Geyman2022Historical2100}. A multi-glacier evaluation is needed to assess the generalisability of the findings presented here.

Third, the prior ensemble is generated from perturbations of a single model's parameters and meteorological forcing, rather than from a multi-model ensemble that considers model structural differences. The question of how large and structurally diverse an ensemble must be to adequately represent prior model structural uncertainty is itself non-trivial \citep{Milinski2020HowBe}, and the single-model approach used here may cause the ensemble spread to underrepresent certain structural uncertainties. This is particularly relevant in the accumulation zone, where the particle degeneracy behaviour identified for snow depth may depend sensitively on the assumed prior distribution and ensemble size. Similarly, the 12-year simulation period, while sufficient for identifying robust mean patterns, provides a limited sample for characterising the tails and thus extremes of interannual variability. Finally, within each glacier zone, spatial heterogeneity in surface properties and topographic shading are not resolved. Real observations at fine spatial resolution, such as ground-based snow depth measurements or high-resolution satellite altimetry \citep{An2020SnowReflectometry, Mazzolini2024Spatio-temporalAltimeter}, may carry additional information not captured in the zone-averaged analysis presented here.

\section{Conclusion}
We conducted large-ensemble synthetic twin experiments on the Kongsvegen glacier, Svalbard, to identify what makes observations effective for data assimilation to improve estimates of glacier surface mass balance. Using the Particle Batch Smoother across three glacier zones, three observation types, two contrasting climatic scenarios, and two observation quality levels, we show that observation quality, quantity, timing, and variable combinations each play distinct roles in determining assimilation performance.

First, high quality observations achieve faster convergence and higher peak improvement across all cases, but sufficient quantities of lower quality observations can compensate for quality deficiencies in all zones except snow depth in the accumulation zone. In the accumulation zone, snow depth assimilation degrades with increasing observation count due to particle degeneracy, a structural failure of the PBS when additional observations are redundant rather than independent in an environment that retains snow. This exception highlights the importance of monitoring posterior weight concentration as a degeneracy diagnostic when these methods are applied to real observations.

Second, optimal observation timing is governed primarily by the structure of the truth trajectory, specifically the timing and sharpness of its seasonal transitions. The best noisy observations consistently coincide with the most rapidly changing phases of the underlying truth signal, namely the sharp albedo decline from snow to ice, the rapid spring snowmelt, and the onset of cooling in the accumulation season. The interannual shift of up to six weeks in optimal albedo timing between early and late melting years directly demonstrates this. The prior parameter distribution is the same in every year, so the interannual shift in optimal timing reflects the year-specific meteorological forcing and the truth trajectories it generates. The S2 scenario reinforces the argument. The prior ensemble is identical in both scenarios, yet the optimal observation timing follows the scenario-specific truth transitions. Adaptive observation scheduling should therefore monitor trajectory transitions in real time, detecting the onset of rapid change in the observed variable, rather than targeting broad seasonal windows defined by prior uncertainty alone.

Third, joint assimilation of independently timed best observations outperforms all individual and same day paired strategies. The benefit arises from temporal diversity. Albedo, snow depth, and surface temperature constrain complementary seasonal phases of the mass and energy balance cycle, and their combined effect is realised only when each is observed at its own optimal time. The combination of all three independently timed observations gives the highest performance, and because the three optimal windows are mutually distinct, it remains stable from year to year, sustaining a high improvement even when a single variable is weakly informative in a given year. Multivariate pairing on the same day sacrifices this temporal complementarity and provides limited or even negative benefit, implying that simultaneous overpasses from multiple sensors can be less valuable for joint assimilation than independently scheduled observations targeting the spread peak of each variable in its own season.

These conclusions provide actionable guidelines for Arctic glacier monitoring. High quality observations should be prioritised where feasible. Observations should be scheduled adaptively by tracking sharp transitions in the variable trajectory rather than following fixed calendars. Observation systems that combine multiple variables should target the seasonal optimum of each variable independently rather than requiring them to coincide in time. Future work should extend this analysis to real rather than synthetic observations and evaluate the sensitivity of these findings to glacier geometry, climate regime, and mass balance model structure across multiple Svalbard and Arctic glaciers.


\section*{Conflict of Interest Statement}

The authors declare that there is no conflict of interest.

\section*{Author Contributions}

Conceptualization was by WC, KA, LSS, and TVS. Data curation was by WC. Formal analysis was by WC, KA, LSS, and TVS. Funding acquisition was by TVS. Methodology was by WC, KA, and TVS. Supervision was by TVS, KA and LSS. Visualization was by WC. Writing- original draft was by WC,  and edited by all co-authors.

\section*{Funding}
 This study was funded by the European Union’s Horizon 2020 research and innovation program under the Marie Skłodowska-Curie action COMPSCI (grant no. 945371), the EEA and Norway initiative HarSval (grant no. UMO-2023/43/7/ST10/00001) and by the European Union's Horizon Europe program through the project LIQUIDICE (grant no. 101184962). Louise S. Schmidt was funded by the Research Council of Norway through the Nansen Legacy project (NFR-276730) and the MAMMAMIA project (NFR-301837). Kristoffer Aalstad acknowledges funding from the ERC-2022-ADG under grant agreement No 101096057 GLACMASS and an ESA CCI Research Fellowship (PATCHES project). 

\section*{Acknowledgments}
The simulations were performed on resources provided by the Department of Geosciences, University of Oslo. 


\section*{Data Availability Statement}
CARRA data was downloaded from the Copernicus Climate Change Service (C3S) Climate Data Store at \\ $https://doi.org/10.24381/cds.d29ad2c6$.
The results are generated using Copernicus Climate Change Service information (2025). Neither the European Commission nor ECMWF is responsible for any use that may be made of the Copernicus information or data it contains. The CryoGrid community model is hosted on Github. The source code is available at \\ $https://github.com/CryoGrid/CryoGridCommunity\_source$. 

\bibliographystyle{Frontiers-Harvard} 
\bibliography{Cubic}

@article{Karner2013ASvalbard,
    title = {{A decade of energy and mass balance investigations on the glacier Kongsvegen, Svalbard}},
    year = {2013},
    journal = {Journal of Geophysical Research Atmospheres},
    author = {Karner, F. and Obleitner, F. and Krismer, T. and Kohler, J. and Greuell, W.},
    number = {10},
    pages = {3986--4000},
    volume = {118},
    doi = {10.1029/2012JD018342},
    issn = {21698996}
}

@article{VanPelt2019A1957-2018,
    title = {{A long-term dataset of climatic mass balance, snow conditions, and runoff in Svalbard (1957-2018)}},
    year = {2019},
    journal = {Cryosphere},
    author = {Van Pelt, Ward and Pohjola, Veijo and Pettersson, Rickard and Marchenko, Sergey and Kohler, Jack and Luks, Bartłomiej and Ove Hagen, Jon and Schuler, Thomas V. and Dunse, Thorben and No{\"{e}}l, Brice and Reijmer, Carleen},
    number = {9},
    pages = {2259--2280},
    volume = {13},
    doi = {10.5194/tc-13-2259-2019},
    issn = {19940424}
}

@techreport{Kaser2003A2003,
    title = {{A Manual for Monitoring the Mass Balance of Mountain Glaciers with Particular Attention to Low Latitude Characteristics}},
    year = {2003},
    author = {Kaser, Georg and Fountain, Andrew and Jansson, Peter},
    institution = {UNESCO, International Hydrological Programme},
    type = {IHP-VI Technical Documents in Hydrology},
    number = {59},
    address = {Paris}
}

@article{Margulis2015AEstimation,
    title = {{A Particle Batch Smoother Approach to Snow Water Equivalent Estimation}},
    year = {2015},
    journal = {Journal of Hydrometeorology},
    author = {Margulis, Steven A. and Girotto, Manuela and Cort{\'{e}}s, Gonzalo and Durand, Michael},
    number = {4},
    month = {8},
    pages = {1752--1772},
    volume = {16},
    publisher = {American Meteorological Society},
    url = {https://journals.ametsoc.org/view/journals/hydr/16/4/jhm-d-14-0177_1.xml},
    doi = {10.1175/JHM-D-14-0177.1},
    issn = {1525-7541}
}

@article{VanPelt2021AcceleratingEnsemble,
    title = {{Accelerating future mass loss of Svalbard glaciers from a multi-model ensemble}},
    year = {2021},
    journal = {Journal of Glaciology},
    author = {Van Pelt, Ward J.J. and Schuler, Thomas V. and Pohjola, Veijo A. and Pettersson, Rickard},
    number = {263},
    pages = {485--499},
    volume = {67},
    doi = {10.1017/jog.2021.2},
    issn = {00221430}
}

@article{Stroeve2005AccuracyMeasurements,
    title = {{Accuracy assessment of the MODIS 16-day albedo product for snow: Comparisons with Greenland in situ measurements}},
    year = {2005},
    journal = {Remote Sensing of Environment},
    author = {Stroeve, Julienne and Box, Jason E. and Gao, Feng and Liang, Shunlin and Nolin, Anne and Schaaf, Crystal},
    number = {1},
    month = {1},
    pages = {46--60},
    volume = {94},
    doi = {10.1016/j.rse.2004.09.001},
    issn = {00344257}
}

@article{Landmann2021AssimilatingFilter,
    title = {{Assimilating near-real-time mass balance stake readings into a model ensemble using a particle filter}},
    year = {2021},
    journal = {Cryosphere},
    author = {Landmann, Johannes Marian and K{\"{u}}nsch, Hans Rudolf and Huss, Matthias and Ogier, Christophe and Kalisch, Markus and Farinotti, Daniel},
    number = {11},
    pages = {5017--5040},
    volume = {15},
    doi = {10.5194/tc-15-5017-2021},
    issn = {19940424}
}

@article{Bertoncini2024AssimilationHydrology,
    title = {{Assimilation of Satellite Albedo to Improve Simulations of Glacier Hydrology}},
    year = {2024},
    volume = {38},
    number = {11},
    journal = {Hydrological Processes},
    author = {Bertoncini, Andr{\'{e}} and Pomeroy, John W.},
    month = {3},
    publisher = {Authorea},
    doi = {10.1002/hyp.15329}
}

@article{Cao2025BayesianExperiments,
    title = {{Bayesian data assimilation on an Arctic glacier: learning from large ensemble twin experiments}},
    year = {2025},
    journal = {Journal of Glaciology},
    author = {Cao, Wenxue and Aalstad, Kristoffer and Schmidt, Louise Steffensen and Westermann, Sebastian and Schuler, Thomas V.},
    month = {11},
    pages = {e121},
    volume = {71},
    publisher = {Cambridge University Press},
    url = {https://www.cambridge.org/core/journals/journal-of-glaciology/article/bayesian-data-assimilation-on-an-arctic-glacier-learning-from-large-ensemble-twin-experiments/819468644553E83EB8E95D6710CF8070},
    doi = {10.1017/jog.2025.10101},
    issn = {17275652}
}

@article{Pramanik2019ComparisonPrecipitation,
    title = {{Comparison of snow accumulation events on two High-Arctic glaciers to model-derived and observed precipitation}},
    year = {2019},
    journal = {Polar Research},
    author = {Pramanik, Ankit and Kohler, Jack and Schuler, Thomas V. and van Pelt, Ward and Cohen, Lana},
    month = {8},
    volume = {38},
    publisher = {Norwegian Polar Institute},
    url = {https://polarresearch.net/index.php/polar/article/view/3364/9369 https://polarresearch.net/index.php/polar/article/view/3364},
    doi = {10.33265/POLAR.V38.3364},
    issn = {1751-8369}
}

@book{Evensen2022DataFundamentals,
    title = {{Data Assimilation Fundamentals}},
    year = {2022},
    author = {Evensen, Geir and Vossepoel, Femke C. and van Leeuwen, Peter Jan},
    series = {Springer Textbooks in Earth Sciences, Geography and Environment},
    publisher = {Springer International Publishing},
    url = {https://link.springer.com/10.1007/978-3-030-96709-3},
    address = {Cham},
    isbn = {978-3-030-96708-6},
    doi = {10.1007/978-3-030-96709-3}
}

@article{Carrassi2018DataPerspectives,
    title = {{Data assimilation in the geosciences: An overview of methods, issues, and perspectives}},
    year = {2018},
    journal = {Wiley Interdisciplinary Reviews: Climate Change},
    author = {Carrassi, Alberto and Bocquet, Marc and Bertino, Laurent and Evensen, Geir},
    number = {5},
    volume = {9},
    doi = {10.1002/wcc.535},
    issn = {17577799},
    arxivId = {1709.02798}
}

@article{Moller2018Differing19002010,
    title = {{Differing Climatic Mass Balance Evolution Across Svalbard Glacier Regions Over 1900–2010}},
    year = {2018},
    journal = {Frontiers in Earth Science},
    author = {M{\"{o}}ller, Marco and Kohler, Jack},
    number = {September},
    pages = {1--20},
    volume = {6},
    doi = {10.3389/feart.2018.00128},
    issn = {22966463}
}

@article{Aalstad2018Ensemble-basedSites,
    title = {{Ensemble-based assimilation of fractional snow-covered area satellite retrievals to estimate the snow distribution at Arctic sites}},
    year = {2018},
    journal = {Cryosphere},
    author = {Aalstad, Kristoffer and Westermann, Sebastian and Schuler, Thomas Vikhamar and Boike, Julia and Bertino, Laurent},
    number = {1},
    pages = {247--270},
    volume = {12},
    doi = {10.5194/tc-12-247-2018},
    issn = {19940424}
}

@article{Paul2017ErrorProject,
    title = {{Error sources and guidelines for quality assessment of glacier area, elevation change, and velocity products derived from satellite data in the Glaciers\_cci project}},
    year = {2017},
    journal = {Remote Sensing of Environment},
    author = {Paul, Frank and Bolch, Tobias and Briggs, Kate and K{\"{a}}{\"{a}}b, Andreas and McMillan, Malcolm and McNabb, Robert and Nagler, Thomas and Nuth, Christopher and Rastner, Philipp and Strozzi, Tazio and Wuite, Jan},
    pages = {1--20},
    url = {http://dx.doi.org/10.1016/j.rse.2017.08.038},
    doi = {10.1016/j.rse.2017.08.038}
}

@article{Alonso-Gonzalez2023ExploringExperiment,
    title = {{Exploring the potential of thermal infrared remote sensing to improve a snowpack model through an observing system simulation experiment}},
    year = {2023},
    journal = {The Cryosphere},
    author = {Alonso-Gonz{\'{a}}lez, Esteban and Gascoin, Simon and Arioli, Sara and Picard, Ghislain},
    pages = {3329--3342},
    volume = {17},
    url = {https://doi.org/10.5194/tc-17-3329-2023},
    doi = {10.5194/tc-17-3329-2023}
}

@article{Schuster2023GlacierStrategies,
    title = {{Glacier projections sensitivity to temperature-index model choices and calibration strategies}},
    year = {2023},
    journal = {Annals of Glaciology},
    author = {Schuster, Lilian and Rounce, David R. and Maussion, Fabien},
    number = {92},
    month = {9},
    pages = {293--308},
    volume = {64},
    publisher = {Cambridge University Press},
    url = {https://www.cambridge.org/core/journals/annals-of-glaciology/article/glacier-projections-sensitivity-to-temperatureindex-model-choices-and-calibration-strategies/EBE19247F4ADC7888EB59EBECB0140B1},
    doi = {10.1017/AOG.2023.57},
    issn = {0260-3055}
}

@article{Zemp2019Global2016,
    title = {{Global glacier mass changes and their contributions to sea-level rise from 1961 to 2016}},
    year = {2019},
    journal = {Nature},
    author = {Zemp, M. and Huss, M. and Thibert, E. and Eckert, N. and McNabb, R. and Huber, J. and Barandun, M. and Machguth, H. and Nussbaumer, S. U. and G{\"{a}}rtner-Roer, I. and Thomson, L. and Paul, F. and Maussion, F. and Kutuzov, S. and Cogley, J. G.},
    number = {7752},
    month = {4},
    pages = {382--386},
    volume = {568},
    publisher = {Nature Publishing Group},
    doi = {10.1038/S41586-019-1071-0},
    issn = {14764687},
    pmid = {30962533}
}

@article{Geyman2022Historical2100,
    title = {{Historical glacier change on Svalbard predicts doubling of mass loss by 2100}},
    year = {2022},
    journal = {Nature},
    author = {Geyman, Emily C. and van Pelt, Ward J. J. and Maloof, Adam C. and Aas, Harald Faste and Kohler, Jack},
    number = {7893},
    pages = {374--379},
    volume = {601},
    publisher = {Springer US},
    doi = {10.1038/s41586-021-04314-4},
    issn = {0028-0836}
}

@article{Milinski2020HowBe,
    title = {{How large does a large ensemble need to be?}},
    year = {2020},
    journal = {Earth System Dynamics},
    author = {Milinski, Sebastian and Maher, Nicola and Olonscheck, Dirk},
    number = {4},
    month = {10},
    pages = {885--901},
    volume = {11},
    publisher = {Copernicus GmbH},
    doi = {10.5194/ESD-11-885-2020},
    issn = {21904987}
}

@article{Kotarba2022ImpactMissions,
    title = {{Impact of the revisit frequency on cloud climatology for CALIPSO, EarthCARE, Aeolus, and ICESat-2 satellite lidar missions}},
    year = {2022},
    journal = {Atmospheric Measurement Techniques},
    author = {Kotarba, Andrzej Z.},
    number = {14},
    month = {7},
    pages = {4307--4322},
    volume = {15},
    publisher = {Copernicus GmbH},
    doi = {10.5194/AMT-15-4307-2022},
    issn = {18678548}
}

@article{Robinson2018ImprovingObservations,
    title = {{Improving Particle Filter Performance by Smoothing Observations}},
    year = {2018},
    journal = {Monthly Weather Review},
    author = {Robinson, Gregor and Grooms, Ian and Kleiber, William},
    number = {8},
    month = {8},
    pages = {2433--2446},
    volume = {146},
    publisher = {American Meteorological Society},
    url = {https://journals.ametsoc.org/view/journals/mwre/146/8/mwr-d-17-0349.1.xml},
    doi = {10.1175/MWR-D-17-0349.1},
    issn = {1520-0493},
    arxivId = {1711.06758}
}

@article{Dumont2012LinkingData,
    title = {{Linking glacier annual mass balance and glacier albedo retrieved from MODIS data}},
    year = {2012},
    journal = {Cryosphere},
    author = {Dumont, M. and Gardelle, J. and Sirguey, P. and Guillot, A. and Six, D. and Rabatel, A. and Arnaud, Y.},
    number = {6},
    pages = {1527--1539},
    volume = {6},
    doi = {10.5194/TC-6-1527-2012},
    issn = {19940416}
}

@article{Berthier2023MeasuringReview,
    title = {{Measuring glacier mass changes from space—a review}},
    year = {2023},
    journal = {Reports on Progress in Physics},
    author = {Berthier, Etienne and Floricioiu, Dana and Gardner, Alex S. and Gourmelen, Noel and Jakob, Livia and Paul, Frank and Treichler, Désirée and Wouters, Bert and Belart, Joaquín M.C. and Dehecq, Amaury and Dussaillant, Ines and Hugonnet, Romain and K{\"{a}}{\"{a}}b, Andreas and Krieger, Lukas and P{\'{a}}lsson, Finnur and Zemp, Michael},
    number = {3},
    month = {2},
    pages = {036801},
    volume = {86},
    publisher = {IOP Publishing},
    doi = {10.1088/1361-6633/ACAF8E},
    issn = {0034-4885},
    pmid = {36596254}
}

@article{Schmidt2023MeltwaterSvalbard,
    title = {{Meltwater runoff and glacier mass balance in the high Arctic: 1991-2022 simulations for Svalbard}},
    year = {2023},
    journal = {EGUsphere},
    author = {Schmidt, Louise Steffensen and Schuler, Thomas V and Thomas, Erin Emily and Westermann, Sebastian},
    pages = {1--32},
    volume = {2023},
    url = {https://doi.org/10.5194/egusphere-2022-1409}
}

@article{Davaze2018MonitoringData,
    title = {{Monitoring glacier albedo as a proxy to derive summer and annual surface mass balances from optical remote-sensing data}},
    year = {2018},
    journal = {Cryosphere},
    author = {Davaze, Lucas and Rabatel, Antoine and Arnaud, Yves and Sirguey, Pascal and Six, Delphine and Letreguilly, Anne and Dumont, Marie},
    number = {1},
    pages = {271--286},
    volume = {12},
    doi = {10.5194/tc-12-271-2018},
    issn = {19940424}
}

@article{Smyth2019ParticleSWE,
    title = {{Particle Filter Data Assimilation of Monthly Snow Depth Observations Improves Estimation of Snow Density and SWE}},
    year = {2019},
    journal = {Water Resources Research},
    author = {Smyth, Eric J. and Raleigh, Mark S. and Small, Eric E.},
    number = {2},
    month = {2},
    pages = {1296--1311},
    volume = {55},
    publisher = {John Wiley {\&} Sons, Ltd},
    url = {https://onlinelibrary.wiley.com/doi/full/10.1029/2018WR023400 https://onlinelibrary.wiley.com/doi/abs/10.1029/2018WR023400 https://agupubs.onlinelibrary.wiley.com/doi/10.1029/2018WR023400},
    doi = {10.1029/2018WR023400},
    issn = {1944-7973}
}

@article{Marzeion2020PartitioningChange,
    title = {{Partitioning the Uncertainty of Ensemble Projections of Global Glacier Mass Change}},
    year = {2020},
    journal = {Earth's Future},
    author = {Marzeion, Ben and Hock, Regine and Anderson, Brian and Bliss, Andrew and Champollion, Nicolas and Fujita, Koji and Huss, Matthias and Immerzeel, Walter W. and Kraaijenbrink, Philip and Malles, Jan Hendrik and Maussion, Fabien and Radi{\'{c}}, Valentina and Rounce, David R. and Sakai, Akiko and Shannon, Sarah and van de Wal, Roderik and Zekollari, Harry},
    number = {7},
    month = {7},
    pages = {e2019EF001470},
    volume = {8},
    publisher = {John Wiley {\&} Sons, Ltd},
    doi = {10.1029/2019EF001470},
    issn = {2328-4277}
}

@article{Navari2021Reanalysis20002014,
    title = {{Reanalysis Surface Mass Balance of the Greenland Ice Sheet Along K-Transect (2000–2014)}},
    year = {2021},
    journal = {Geophysical Research Letters},
    author = {Navari, Mahdi and Margulis, Steven A. and Tedesco, Marco and Fettweis, Xavier and van de Wal, Roderik S.W.},
    number = {17},
    month = {9},
    pages = {e2021GL094602},
    volume = {48},
    publisher = {John Wiley {\&} Sons, Ltd},
    url = {https://onlinelibrary.wiley.com/doi/full/10.1029/2021GL094602 https://onlinelibrary.wiley.com/doi/abs/10.1029/2021GL094602 https://agupubs.onlinelibrary.wiley.com/doi/10.1029/2021GL094602},
    doi = {10.1029/2021GL094602},
    issn = {1944-8007}
}

@article{Schuler2020ReconcilingBalance,
    title = {{Reconciling Svalbard Glacier Mass Balance}},
    year = {2020},
    journal = {Frontiers in Earth Science},
    author = {Schuler, Thomas V. and Kohler, Jack and Elagina, Nelly and Hagen, Jon Ove M. and Hodson, Andrew J. and Jania, Jacek A. and K{\"{a}}{\"{a}}b, Andreas M. and Luks, Bartłomiej and Ma{\l}ecki, Jakub and Moholdt, Geir and Pohjola, Veijo A. and Sobota, Ireneusz and Van Pelt, Ward J.J.},
    number = {May},
    pages = {1--16},
    volume = {8},
    doi = {10.3389/feart.2020.00156},
    issn = {22966463}
}

@article{stby2014SevereSvalbard,
    title = {{Severe cloud contamination of MODIS Land Surface Temperatures over an Arctic ice cap, Svalbard}},
    year = {2014},
    journal = {Remote Sensing of Environment},
    author = {{\O}stby, Torbjørn I. and Schuler, Thomas V. and Westermann, Sebastian},
    month = {2},
    pages = {95--102},
    volume = {142},
    publisher = {Elsevier},
    doi = {10.1016/J.RSE.2013.11.005},
    issn = {0034-4257}
}

@article{An2020SnowReflectometry,
    title = {{Snow Depth Variations in Svalbard Derived from GNSS Interferometric Reflectometry}},
    year = {2020},
    journal = {Remote Sensing 2020, Vol. 12, Page 3352},
    author = {An, Jiachun and Deng, Pan and Zhang, Baojun and Liu, Jingbin and Ai, Songtao and Wang, Zemin and Yu, Qiuze},
    number = {20},
    month = {10},
    pages = {3352},
    volume = {12},
    publisher = {Multidisciplinary Digital Publishing Institute},
    url = {https://www.mdpi.com/2072-4292/12/20/3352/htm https://www.mdpi.com/2072-4292/12/20/3352},
    doi = {10.3390/RS12203352},
    issn = {2072-4292}
}

@article{Mazzolini2024Spatio-temporalAltimeter,
    title = {{Spatio-temporal snow data assimilation with the ICESat-2 laser altimeter}},
    year = {2025},
    journal = {The Cryosphere},
    author = {Mazzolini, Marco and Aalstad, Kristoffer and Alonso-Gonz{\'{a}}lez, Esteban and Westermann, Sebastian and Treichler, Désirée},
    number = {9},
    pages = {3831--3848},
    volume = {19},
    doi = {10.5194/tc-19-3831-2025}
}

@article{Vionnet2012TheV7.2,
    title = {{The detailed snowpack scheme Crocus and its implementation in SURFEX v7.2}},
    year = {2012},
    journal = {Geoscientific Model Development},
    author = {Vionnet, V. and Brun, E. and Morin, S. and Boone, A. and Faroux, S. and Le Moigne, P. and Martin, E. and Willemet, J. M.},
    number = {3},
    pages = {773--791},
    volume = {5},
    doi = {10.5194/GMD-5-773-2012},
    issn = {1991959X}
}

@article{Alonso-Gonzalez2022TheV1.0,
    title = {{The Multiple Snow Data Assimilation System (MuSA v1.0)}},
    year = {2022},
    journal = {Geoscientific Model Development},
    author = {Alonso-Gonz{\'{a}}lez, Esteban and Aalstad, Kristoffer and Baba, Mohamed Wassim and Revuelto, Jesús and L{\'{o}}pez-Moreno, Juan Ignacio and Fiddes, Joel and Essery, Richard and Gascoin, Simon},
    number = {24},
    month = {12},
    pages = {9127--9155},
    volume = {15},
    publisher = {Copernicus Publications},
    doi = {10.5194/GMD-15-9127-2022},
    issn = {19919603}
}

@article{Ye2024UnveilingGlaciers,
    title = {{Unveiling Glacier Mass Balance: Albedo Aggregation Insights for Austrian and Norwegian Glaciers}},
    year = {2024},
    journal = {Remote Sensing 2024, Vol. 16, Page 1914},
    author = {Ye, Fan and Cheng, Qing and Hao, Weifeng and Hu, Anxun and Liang, Dong},
    number = {11},
    month = {5},
    pages = {1914},
    volume = {16},
    publisher = {Multidisciplinary Digital Publishing Institute},
    doi = {10.3390/RS16111914},
    issn = {2072-4292}
}

@article{Dumont2012VariationalGlacier,
    title = {{Variational assimilation of albedo in a snowpack model and reconstruction of the spatial mass-balance distribution of an alpine glacier}},
    year = {2012},
    journal = {Journal of Glaciology},
    author = {Dumont, Marie and Durand, Yves and Arnaud, Yves and Six, Delphine},
    number = {207},
    month = {2},
    pages = {151--164},
    volume = {58},
    publisher = {Cambridge University Press},
    doi = {10.3189/2012JOG11J163},
    issn = {0022-1430}
}

@article{Westermann2023CryoGridCommunity,
    title = {{The CryoGrid community model (version 1.0) - a multi-physics toolbox for climate-driven simulations in the terrestrial cryosphere}},
    year = {2023},
    journal = {Geoscientific Model Development},
    author = {Westermann, Sebastian and Ingeman-Nielsen, Thomas and Scheer, Johanna and Aalstad, Kristoffer and Aga, Juditha and Chaudhary, Nitin and Etzelm{\"{u}}ller, Bernd and Filhol, Simon and K{\"{a}}{\"{a}}b, Andreas and Renette, Cas and Schmidt, Louise Steffensen and Schuler, Thomas Vikhamar and Zweigel, Robin B. and Martin, L{\'{e}}o and Morard, Sarah and Ben-Asher, Matan and Angelopoulos, Michael and Boike, Julia and Groenke, Brian and Miesner, Frederieke and Nitzbon, Jan and Overduin, Paul and Stuenzi, Simone M. and Langer, Moritz},
    number = {9},
    pages = {2607--2647},
    volume = {16},
    doi = {10.5194/gmd-16-2607-2023}
}

@misc{C3S2024CARRA,
    title = {{Arctic regional reanalysis on single levels from 1991 to present}},
    year = {2024},
    author = {{Copernicus Climate Change Service}},
    howpublished = {Copernicus Climate Change Service (C3S) Climate Data Store (CDS)},
    doi = {10.24381/cds.713858f6}
}

@article{GlaMBIE2025CommunityChanges,
    title = {{Community estimate of global glacier mass changes from 2000 to 2023}},
    year = {2025},
    journal = {Nature},
    author = {{The GlaMBIE Team}},
    number = {8054},
    month = {2},
    pages = {382--388},
    volume = {639},
    doi = {10.1038/s41586-024-08545-z},
    url = {https://www.nature.com/articles/s41586-024-08545-z}
}

@article{Montzka2012MultivariateReview,
    title = {{Multivariate and Multiscale Data Assimilation in Terrestrial Systems: A Review}},
    year = {2012},
    journal = {Sensors},
    author = {Montzka, Carsten and Pauwels, Valentijn R. N. and Hendricks Franssen, Harrie-Jan and Han, Xujun and Vereecken, Harry},
    number = {12},
    pages = {16291--16333},
    volume = {12},
    doi = {10.3390/s121216291}
}

@article{Largeron2020TowardReview,
    title = {{Toward Snow Cover Estimation in Mountainous Areas Using Modern Data Assimilation Methods: A Review}},
    year = {2020},
    journal = {Frontiers in Earth Science},
    author = {Largeron, Chlo{\'{e}} and Dumont, Marie and Morin, Samuel and Boone, Aaron and Lafaysse, Matthieu and Metref, Sammy and Cosme, Emmanuel and Jonas, Tobias and Winstral, Adam and Margulis, Steven A.},
    pages = {325},
    volume = {8},
    doi = {10.3389/feart.2020.00325}
}

@article{Fowler2012MeasuresAssimilation,
    title = {{Measures of observation impact in non-Gaussian data assimilation}},
    year = {2012},
    journal = {Tellus A: Dynamic Meteorology and Oceanography},
    author = {Fowler, Alison and {Van Leeuwen}, Peter Jan},
    pages = {17192},
    volume = {64},
    doi = {10.3402/tellusa.v64i0.17192}
}

@misc{Kohler2021SvalbardInventory,
    title = {{Svalbard glacier inventory based on Sentinel-2 imagery from summer 2020}},
    year = {2021},
    author = {Kohler, Jack and Lith, Aniek and Moholdt, Geir},
    howpublished = {Norwegian Polar Institute},
    doi = {10.21334/npolar.2021.1b8631bf}
}

@article{Moller2017ModelingHiRSvaC,
    title = {{Modeling glacier-surface albedo across Svalbard for the 1979-2015 period: The HiRSvaC500-$\alpha$ data set}},
    year = {2017},
    journal = {Journal of Advances in Modeling Earth Systems},
    author = {M{\"{o}}ller, Marco and M{\"{o}}ller, Rebecca},
    number = {1},
    pages = {404--422},
    volume = {9},
    doi = {10.1002/2016MS000752}
}

@article{Aalstad2020,
    title = {{Evaluating satellite retrieved fractional snow-covered area at a high-Arctic site using terrestrial photography}},
    year = {2020},
    journal = {Remote Sensing of Environment},
    author = {Aalstad, Kristoffer and Westermann, Sebastian and Bertino, Laurent},
    pages = {111618},
    volume = {239},
    doi = {10.1016/j.rse.2019.111618}
}

@article{Margulis2016,
    title = {{A Landsat-era Sierra Nevada snow reanalysis (1985-2015)}},
    year = {2016},
    journal = {Journal of Hydrometeorology},
    author = {Margulis, Steven A. and Cort{\'{e}}s, Gonzalo and Girotto, Manuela and Durand, Michael},
    number = {4},
    pages = {1203--1221},
    volume = {17},
    doi = {10.1175/JHM-D-15-0177.1}
}

@book{Chopin2020,
    title = {{An Introduction to Sequential Monte Carlo}},
    year = {2020},
    author = {Chopin, Nicolas and Papaspiliopoulos, Omiros},
    publisher = {Springer},
    doi = {10.1007/978-3-030-47845-2}
}

@article{Herrmann2025,
    title = {{A Kalman filter-based framework for assimilating remote sensing observations into a surface mass balance model}},
    year = {2025},
    journal = {Annals of Glaciology},
    author = {Herrmann, Oskar and Groos, Alexander R. and Tabone, Ilaria and Jouvet, Guillaume and F{\"{u}}rst, Johannes J.},
    pages = {e23},
    volume = {66},
    doi = {10.1017/aog.2025.10020}
}

@article{Smyth2020ImprovingTiming,
    title = {{Improving SWE Estimation With Data Assimilation: The Influence of Snow Depth Observation Timing and Uncertainty}},
    year = {2020},
    journal = {Water Resources Research},
    author = {Smyth, Eric J. and Raleigh, Mark S. and Small, Eric E.},
    number = {5},
    pages = {e2019WR026853},
    volume = {56},
    doi = {10.1029/2019WR026853}
}

@article{Guidicelli2024,
    title = {{A combined data assimilation and deep learning approach for continuous spatio-temporal SWE reconstruction from sparse ground tracks}},
    year = {2024},
    journal = {Journal of Hydrology X},
    author = {Guidicelli, Matteo and Aalstad, Kristoffer and Treichler, D{\'{e}}sir{\'{e}}e and Salzmann, Nadine},
    pages = {100190},
    volume = {25},
    doi = {10.1016/j.hydroa.2024.100190}
}

@article{Aalstad2026,
    title = {{Evolving beyond collapse: An adaptive particle batch smoother for cryospheric data assimilation}},
    year = {2026},
    journal = {arXiv preprint},
    author = {Aalstad, Kristoffer and Alonso-Gonz{\'{a}}lez, Esteban and Pirk, Norbert and Westermann, Sebastian and Willmes, Christian and Yang, Ruitang},
    doi = {10.48550/arXiv.2601.20049}
}

@article{Hersbach2000,
    title = {{Decomposition of the Continuous Ranked Probability Score for Ensemble Prediction Systems}},
    year = {2000},
    journal = {Weather and Forecasting},
    author = {Hersbach, Hans},
    number = {5},
    pages = {559--570},
    volume = {15},
    doi = {10.1175/1520-0434(2000)015<0559:DOTCRP>2.0.CO;2}
}

@article{Gneiting2005,
    title = {{Calibrated Probabilistic Forecasting Using Ensemble Model Output Statistics and Minimum CRPS Estimation}},
    year = {2005},
    journal = {Monthly Weather Review},
    author = {Gneiting, Tilmann and Raftery, Adrian E. and {Westveld III}, Anton H. and Goldman, Tom},
    number = {5},
    pages = {1098--1118},
    volume = {133},
    doi = {10.1175/MWR2904.1}
}

@article{vanHove2026,
    title = {{Actively inferring methane sources with drones}},
    year = {2026},
    journal = {Environmental Data Science},
    author = {van Hove, A. and Aalstad, K. and Pirk, N.},
    pages = {1--29},
    volume = {5},
    doi = {10.1017/eds.2026.10029}
}

@article{Morzfeld2017,
    title = {{What the collapse of the ensemble Kalman filter tells us about particle filters}},
    year = {2017},
    journal = {Tellus A: Dynamic Meteorology and Oceanography},
    author = {Morzfeld, M. and Hodyss, D. and Snyder, C.},
    pages = {1--14},
    volume = {69},
    doi = {10.1080/16000870.2017.1283809}
}

@article{Teweldebrhan2019,
    title = {{Improving the Informational Value of MODIS Fractional Snow Cover Area Using Fuzzy Logic Based Ensemble Smoother Data Assimilation Frameworks}},
    year = {2019},
    journal = {Remote Sensing},
    author = {Teweldebrhan, Aynom T. and Burkhart, John F. and Schuler, Thomas V. and Xu, Chong-Yu},
    number = {1},
    volume = {11},
    doi = {10.3390/rs11010028}
}

@article{Lindley1956,
    title = {{On a Measure of the Information Provided by an Experiment}},
    year = {1956},
    journal = {The Annals of Mathematical Statistics},
    author = {Lindley, D. V.},
    number = {4},
    pages = {986--1005},
    volume = {27},
    doi = {10.1214/aoms/1177728069}
}

@incollection{Masutani2010,
    title = {Observing System Simulation Experiments},
    year = {2010},
    author = {Masutani, Michiko and Schlatter, Thomas W. and Errico, Ronald M. and Stoffelen, Ad and Andersson, Erik and Lahoz, William and Woollen, John S. and Emmitt, G. David and Riish{\o}jgaard, Lars-Peter and Lord, Stephen J.},
    editor = {Lahoz, William and Khattatov, Boris and Menard, Richard},
    booktitle = {Data Assimilation: Making Sense of Observations},
    publisher = {Springer Berlin Heidelberg},
    address = {Berlin, Heidelberg},
    pages = {647--679},
    doi = {10.1007/978-3-540-74703-1_24}
}
\end{document}